\documentstyle[11pt]{article}

\def\bi{\bibitem}
\def\be{\begin{equation}}
\def\ee{\end{equation}}

\def\beql{\arraycolsep .5mm \begin{eqnarray}}
\def\eeql{\end{eqnarray}}
\def\zeile{\nonumber \\[2mm] }   

\def\ea{\end{array}}
\def\la{\label}
\def\ci{\cite}

\def\Ref#1{(\ref{#1})}

\def\R{{\bf R}}
\def\r{\rho}

\def\s{\sigma}

\def\t{\tau}

\def\a{\alpha}
\def\b{\beta}
\def\g{\gamma}
\def\d{\delta}
\def\eps{\epsilon}
\def\veps{\varepsilon}

\def\o{\omega}

\def\ve{\varphi}
\def\x{\xi}

\def\th{\theta}

\def\m{\mu}
\def\n{\nu}

\def\G{\Gamma}
\def\O{\Omega}
\def\T{\Theta}
\def\SO{{\rm SO}}
\def\ISO{{\rm ISO}}
\def\cL{{\cal L}}
\def\p{\partial}
\def\frc#1#2{{\textstyle \frac{#1}{#2}}}

\newcommand{\mkreis}[1]{\mathaccent23 #1}
\def\E#1#2{ E_{#1}^{\;\; #2} }
\def\H#1#2{ H_{#1}^{\;\; #2} }
\def\e#1#2{ e_{#1}^{\;\; #2} }
\def\EN{{\mkreis{E}}}
\def\Enull#1#2{ {{\mkreis{E}}_{#1}}^{\;\; #2} }

\def\Gnull{ {\mkreis{G}} }

\def\Pnull{ {\mkreis{P}} }

\def\gnull{g}
\def\Gamnull{ {\mkreis{\Gamma}} }
\def\onull{{\mkreis{\omega}} }
\def\Onull{{\mkreis{\Omega}} }

\def\Dti{{\tilde{D}}}
\def\Gamti{{\tilde{\Gamma}}}
\def\Rti{{\tilde{R}}}
\def\oti{{\tilde{\omega}}}
\def\Tti{{\tilde{T}}}
\def\Oti{{\tilde{\Omega}}}

\def\Pti{{\tilde{P}}}

\def\hti{{\tilde{h}}}

\def\Nti{{\tilde{N}}}
\def\cA{{\cal A}}
\def\cF{{\cal F}}

\begin{document}
\begin{flushright}    hep-th/9412002  \\
                       LPTENS 94/21 \\
                      DESY 94-156        \end{flushright}
\begin{center}
\LARGE{{\bf Null-Killing Vector Dimensional Reduction and Galilean
Geometrodynamics}\footnote{This work is dedicated to the memory
of Feza G\"ursey.}}  \\
\vspace*{1cm}
\large{B. Julia}  \\
\medskip
Laboratoire de Physique Th\'eorique de l'ENS \\
24, rue Lhomond  75231 Paris Cedex 05, France

\vskip0.5cm
\large{H. Nicolai}  \\
\medskip
II. Institute for Theoretical Physics, Hamburg University, \\
Luruper Chaussee 149,  Hamburg 22761, Germany
\end{center}
\vskip1.0cm
{\bf ABSTRACT.}

The solutions of Einstein's equations admitting
one non-null Killing vector field are best studied with the
projection formalism of Geroch. When the Killing vector
is lightlike, the projection onto the orbit space still
exists and one expects a covariant theory with degenerate
contravariant metric to appear, its geometry is presented here.
Despite the complications of indecomposable
representations of the local Euclidean subgroup, one obtains
an absolute time and a canonical, Galilean and so-called
Newtonian, torsionless connection.
The quasi-Maxwell field (Kaluza Klein one-form) that appears in the
dimensional reduction is a non-separable part of this affine
connection, in contrast to the reduction with a non-null Killing vector.
One may define the Kaluza Klein scalar (dilaton) together with
the absolute time coordinate after having imposed one of the
equations of motion in order to prevent the emergence of torsion.
We present a detailed analysis of the dimensional reduction using moving
frames, we derive the complete equations of motion and
propose an action whose variation gives rise to all but one of
them. Hidden symmetries are shown to act on the space of solutions.

\newpage

\section{Introduction}

In this paper we study the dimensional reduction of Einstein's
theory from $d+1$ dimensions to $d$ dimensions with a
{\sl null Killing vector}. In contrast to the usual Kaluza Klein
reduction of Einstein's theory, on which an ample literature
exists \ci{KK, Ger}, this case has not received much attention until now.
It is nonetheless important for several reasons.
First of all, the analogs of Ehler's group and more
general hidden symmetries known to arise in the dimensional
reduction with a non-null Killing vector have not yet been studied.
A knowledge of such hidden symmetries would facilitate the
analysis of gravitational ``wave" solutions in general
relativity (of which the so-called $pp$-waves are special
examples \ci{Bri}); we note that such exact solutions have recently attracted
renewed interest in connection with string theory \ci{Kall}. The same can
be said of the infinite dimensional symmetries, such as the
Geroch group \ci{Ger2}, which arise in the dimensional reduction of
gravity, supergravities or superstring theories to two dimensions,
and their possible extensions \ci{Julia}. In fact, when dealing with
such generalizations, the question of null-Killing
vectors {\sl must} be addressed \ci{Nic}. Finally the use of moving frames
sheds some new light on the subtleties of Galilean invariant theories with
coordinate reparametrization invariance which are potentially relevant in
the theory of continuous media, in the study of nonrelativistic limits
and possibly in the study of light cone frame dynamics. It is well known that
the (Wigner) little group of a null vector is the Euclidean group, we shall
discuss its gauge realization in curved spacetime.

The non-null reduction of Einstein's gravity from $d+1$
dimensions to $d$ dimensions is well known to give rise to
gravity coupled to a Maxwell and a scalar fields in $d$ dimensions.
The reduced theory is economically described in the moving
frame formalism by use of an orthonormal frame (vielbein)
\be
\E{M}{A} = \pmatrix{ S^{-1/(d-2)} \e{m}{a} & S A_m \cr
                    0   & S \cr } \la{vielbein}
\ee
where $\e{m}{a}$ characterizes the $d$-dimensional gravitational
background, and $A_m$ and $S$ are the Maxwell and scalar matter fields
living on this background. The appropriate Weyl rescalings
of $\e{m}{a}$ and $A_m$ are included so as to obtain the canonically
normalized Einstein Lagrangian in $d$
dimensions and the proper identification
of the Maxwell gauge transformations. The special triangular form
of $\E{M}{A}$ in \Ref{vielbein} is arrived at
by making partial use of local Lorentz invariance.
The Killing vector corresponding to this reduction is taken to have
components $\xi^M=(0,...,0,1)$. Labeling the last coordinate, on which
the dimensional reduction is performed, by $v$ (see below for a
comprehensive summary of our conventions and notation), we thus have
\be
\xi \equiv \xi^M \p_M = \p_v \equiv \frac{\p}{\p v}    \la{Killing1}
\ee
Since the metric is $G_{MN} =
\E{M}{A} E_{NA}$, it is easy to see that, with the form \Ref{vielbein}
of the vielbein, $\xi^2 \equiv \xi^M \xi_M = S^2$ vanishes only for a
degenerate metric. Therefore the Killing direction is assumed to be
non-null but this does not restrict small variations of the metric.
Consequently the above choice of frame is unsuitable to study the
reduction with a null Killing vector, for which $\xi^2 =0$.

The special nature of the dimensional reduction with a null Killing
vector is also evident when the metric is written in the
following form, valid for arbitrary $\xi^2$,
\be
G_{MN} = \pmatrix{ G_{mn} & \xi_m \cr
                   \xi_n   &  \xi^2 \cr}    \la{metric}
\ee
whose inverse we parametrize as follows
\be
G^{MN} = \pmatrix{ h^{mn} & N^m  \cr
                   N^n & N^v \cr }    \la{invmetric}
\ee
Setting $\xi^2 =0$ here corresponds to freezing one component
of the metric to zero (i.e. $G_{vv}=0$), we would therefore
loose one equation of motion (roughly speaking $R^{vv}=0$) if we were to
stick to an action principle. For this reason, we will mostly work
with the equations of motion, although a candidate action will be
presented in section 6.2. For $\xi^2 =0$, the contravariant
metric $h^{mn}$ is degenerate, because then $h^{mn} \xi_n =0$.
This is the reason why in this case we end up with a
``generally covariant" Galilean theory in $d$ dimensions. As shown in
\ci{We, Tr, Do, Ku, Eh2}, such theories possess a pair
of covariantly constant tensors $(h^{mn} , u_m)$, where
the contravariant metric
$h^{mn}$ is degenerate and $u_m$ is its zero eigenvector (suitably
normalized). $h^{mn}$ is essentially the direct image
of $G^{MN}$ on the orbit space of the null Killing flow. Strictly speaking
the generalized Galilean structure we will discover is the kinematic part
of the
complicated set of assumptions needed to formulate pure Newtonian gravity.
In our case the $d$-dimensional geometry will be simpler and disentangled
from the equations
of motion but it will describe gravity plus "electromagnetism" as we will see.

There is a second reason for manipulating the equations of motion
rather than some action, it is the property of the orbits of
the Killing field to be twistless (this is the technical term if
$d=3$), provided another one of the classical equations of motion
($R_{vv}=0$, to be precise) is satisfied. We shall prove that
this property, which is
more generally called ``normality'' of the null Killing vector
field, holds {\sl in any dimension}. In other words we have
\be
\x_M = W \p_M u . \la{normality}
\ee
It is this consequence of the classical equation of motion: $R_{vv}=0$ that
will allow us to construct a torsion-free
connection in $d$ dimensions. In previous work \ci{Du}, the
vanishing of the twist followed from a stronger assumption, namely
the existence of a ``Bargmann" structure or its consequence: the covariant
constancy of the (null) Killing vector;
this restriction is not needed here. We will see that in our approach the
absence of torsion implies the existence of a coordinate $u$,
that will be interpreted as absolute time. The latter is indispensable
in the context of Galilean covariant theories;
although the so-called Newton-Cartan theories of Galilean
relativity are in principle compatible with a nonvanishing
torsion, torsion has never been required until now. Here, we
will see that non-trivial torsion can be eliminated
by transmuting it into the scalar
field $W$ (dilaton) as a consequence of \Ref{normality}. Technically the
scalar can even be made to appear at the same place as in the nonnull case
thanks to the
existence of a Lorentz boost symmetry in $d+1$ dimensions. This scalar
field emerges in our work as a genuine local degree of freedom; but as it
is only defined up to a constant, it will appear solely through its
logarithmic derivatives in the final equations of motion.
When it is replaced by a constant, our results are
compatible with those of \ci{Du}.

The Kaluza Klein one-form, on the other hand, will be shown to
disappear inside the Galilean connection. Contrary to
the non-null case it does not exist on the Killing
orbit space! In fact there is no canonical abelian connection, and one
cannot reinterpret the changes of section of this fibration
of the $(d+1)$-dimensional manifold as Maxwell gauge transformations
as in the nonnull case. More precisely there are frame dependent quasi-Maxwell
fields that will appear in the intermediate steps of our discussion. In order
to emphasize the difference with the usual situation,
we shall call the changes of section $\veps$-{\sl gauge
transformations}. This surprise is compatible with the well-known
fact that the Lorentz force exerted by the Maxwell field on a test
particle moving in this geometry can be reinterpreted as
a Galilean gravitational effect. A generalized Coriolis force corresponds to
the magnetic term and the electric field to the Newtonian one up to the
$e/m$ ratio.

We shall use the following conventions throughout this paper:
capital letters $M,N,..$
and $A,B,...$ will denote curved and flat indices, respectively,
in $d+1$ dimensions. In the reduction to $d$ dimensions, the curved
indices are split as $M=(m,v)$, where $m=1,...,d$ and $v$ is the
index for the coordinate $v$ along the Killing orbits, so $\p_v$
is always a null vector. Flat (Lorentz) indices
$A,B,...$ are split into transverse indices
$a,b,...=1,...,d-1$ and longitudinal indices $(+,-)$, such that
$+$ is the flat homolog of the index $v$, and the
tangent space metric has the light cone frame form:
\be
\eta_{ab} = \d_{ab} \;\;\; , \;\;\; \eta_{+-} =1  \la{flatmetric}
\ee
with all other components vanishing.
When dealing with Kaluza Klein
matter we shall also need to make use
of {\sl intermediate} indices $\a ,\b $ in $(d+1)$
dimensions; they correspond to another anholonomic frame and decompose
as $\a = (\mu , \ve)$, where
$\mu =1,...,d$ and $\ve$ is the intermediate homolog of
$v$ and the flat index $+$. The intermediate frame allows an
$\veps$-invariant but Lorentz dependent separation of {\sl background} and
{\sl matter} fields.

We now summarize the contents of this paper.
First we shall show quite generally that a null Killing
vector is twist-free provided one of Einstein's equations is
satisfied. Frames and symmetries are introduced in the next section.
Symmetries include $d$-dimensional diffeomorphisms, the one parameter
$\veps$-gauge invariance and local Lorentz invariance
partially fixed to an $\ISO(d-1)$ local subgroup.
On a first perusal, readers may then jump to section 5.3 to find
a quick derivation of an affine connection in $d$ dimensions.
However a deeper understanding will come from returning to
section 4 where we set up a $d$-bein formalism to study the case
of pure {\sl background} geometry on the space of Killing orbits and discuss
its most general connection. We reformulate these results
in $(d+1)$-covariant form after having established the correspondence with
earlier work on covariant Newtonian theories. The splitting of matter and
background gravitational
field is not independent of our choice of frame, but it permits
a manifestly $\veps$-gauge invariant treatment.
The geometry with matter is discussed
in section 5. There we construct in particular
the fully covariant $d$-dimensional affine connection; this requires
a modified version of the usual Weyl rescaling, which is here
forced upon us by the symmetry and not by a canonical normalization of
the action as in the non-null case. Further peculiarities of
Galilean physics are analyzed, in particular the
non-separability of the electromagnetic field. Alternative methods permit
the rederivation of the connection and a systematic study of tensor
fields.
The equations for the scalar field, the metric and the connection
are given in section 6. As far as the hidden duality group is concerned,
a kind of contraction of Ehlers' $SO(2)$ action
still exists as suggested by \ci{Ger}. We shall mention that its action
reduces to an $\veps$-gauge transformation in the special case of
$pp$-waves but it acts less
trivially on the van Stockum solutions or their generalizations.
Finally we discuss the
possibility to find an action principle in $d$ dimensions.
We defer the introduction of true extra matter fields to our next paper.
Let us also note that we shall work locally
and postpone temporarily topological and global questions.

\section{Properties of a null Killing field.}

Let us consider a pseudo-Riemannian manifold admitting a null Killing
vector field $\xi^M$: $D_{(M}\xi_{N)}=0$. It is clearly geodesic, i.e.
nonaccelerating
($\x^N D_N\x_M = 0$), divergenceless (i.e. $D_M\xi^M =0$) and
already affinely normalized; it is also by definition
``shearfree'' ($D_{(M}\xi_{N)}=0$). We shall now derive a very important
general consequence of Einstein's equations for classical solutions
admitting a null Killing vector. Contracting the Ricci tensor $R_{MN}$
with $\xi^M \xi^N$, we obtain
\be
0=R_{MN}\x^M\x^N=\x^MG^{PQ}[D_M,D_P]\x_Q   \la{Rmnxi}
\ee
Using the Killing equation and the property
\be
\x^M D_N\x_M = 0
\ee
(which holds for any null vector), we find
\be
D_M\x_N D^M\x^N = 0 = \x_{MN} \x^{MN}  \la{ximn2}
\ee
where $\x_{MN}:=D_M\x_N-D_N\x_M$. We shall keep that tensor convention
of adding one lower index for the exterior derivative in this paper.
We next observe that, due to the equality $\xi^N D_N \xi_M =0$
and the Killing property, we have $\x^M\x_{MN}=0$.

Let us now consider first the case $d=3$. Squaring the expression
$\eps^{MNPQ} \xi_{MN} \xi_P V_Q$, where $V_M$ is an arbitrary
vector, it is easy to see that all contractions vanish, and therefore
\be
\epsilon_{MNPQ}\x^N\x^{PQ} = 0,   \la{hyperso}
\ee
since $V_M$ was arbitrary.
We will refer to this property as ``normality of the Killing vector''
(and not use the word ``hypersurface-orthogonality'' for
esthetic reasons). It implies the result \Ref{normality}
stated in the introduction. By Frobenius' theorem, the null
planes orthogonal to (and containing) the Killing vector
form an integrable system tangent to $d$-manifolds. Owing to
\Ref{normality}, we can define a new
coordinate $u$, which is in some sense the curved analog of
the flat minus coordinate. $u$ is an absolute affine time of the gravitational
solution that replaces in a way the proper time of cosmological matter in a
Friedmann universe. Note however that the vector field $\p/\p u$ has not been
defined yet, it depends on the choice of the other coordinates and is in
general {\sl non-null.}
The function $W$ is an integrating factor; the special case
when it is constant corresponds
to the so-called $pp$-waves \ci{Bri}, it is also the case considered in
\ci{Du, Du2}. Let us stress that what follows holds
irrespective of the assumptions made by these authors.
Observe that $G^{MN}\p_N u$ is also a null geodesic vector field
affinely parametrized and hence $ W$ is constant along
each null geodesic:
\be
\xi^M \p_M W = 0 . \la{EqW}
\ee

Actually, the normality property proved above holds not just in
four but in any number of dimensions. To see this, we can either
repeat the above argument with a set of mutually orthogonal
vectors $V^{(i)}_M$, or otherwise rephrase it with flat indices.
Let us note that the proof of normality in
dimension greater than four relies on the Minkowskian signature of the
metric; in four
or three dimensions, however, the existence of a single time
direction is not required.

We may mention that the above argument can also be turned around (see for
example \ci{Kra}): if the Killing
vector obeys \Ref{hyperso} and is null, the energy momentum
tensor is constrained to obey $\xi^M \xi^N T_{MN} =0$ by
Einstein's equations, regardless of the specific kind of matter
that is coupled to gravity in $d+1=4$ dimensions.
For completeness let us also recall that in $4$ dimensions the vacuum
solutions of Einstein's equations admitting a geodesic non-expanding
non-twisting null congruence form the Kundt class \ci{Kun}. They are all
algebraically special. As they are shearfree they are precisely our
solutions admitting a null Killing vector.

\section{Generalities, symmetries and frames}
There will be two groups of symmetries beyond $d$-dimensional
diffeomorphisms: the Maxwell
type invariance ($\veps$-invariance) corresponding to the arbitrary
choice of sections through the
Killing orbits (the transformation rules are given in eq.\Ref{SEC}), and the
change of transverse vector $n^m$ or more generally the local Lorentz subgroup
$\ISO(d-1)\times \R$ preserving our partial choice of gauge; we shall speak
somewhat abusively of Lorentz invariance for the latter
invariance. Contrary to the non-null case, the
orthogonal space to the null Killing vector
contains the Killing vector itself; thus although it is of
codimension one, it does not provide the rest of a basis for the full
space. So there is no {\sl canonical} abelian
connection, it would depend on the choice of an extra vector field via, as
we shall see, the choice of moving frame.
We shall first list the formulas for the frames to be used and then
motivate our choices by group theoretical arguments.

General covariance in
$d$ dimensions and $\veps$-invariance are preserved by the choice
of what we call an {\sl intermediate moving frame}. This is a
particular choice of Cartan anholonomic frame field that
is only partly null like the light cone
Lorentz frame defined above. But it will allow a convenient and
$\veps$-covariant separation of the gravitational background from the
Maxwell and scalar matter fields. In the non-null case
\Ref{vielbein}, this frame implements a fully covariant separation of
$\e{m}{a}$ from the bona fide matter fields $(A_m ,S)$.
Let us insist however on the unusual fact that here this
separation of a Maxwell field from  the {\sl gravitational}
field is Lorentz-noncovariant and hence a temporary {\sl artefact} of
our discussion; therefore our designation of both $S$ and $A_m$ as
{\sl matter fields} involves some abuse of language.

In fact the {\sl main difficulty} will be to reconcile
$\veps$ and $\ISO(d-1)$ invariances and to define the appropriate tensor
calculus. The idea is to implement successively these invariances, firstly
in this order beginning in section 4 and then, with hindsight, in the
reverse order in section 5.3.

\subsection{Frames}
Accordingly, we represent the full moving
frame as a product of a background vielbein $\Enull{\a}{A}$ and
an intermediate frame $\H{M}{\a}$ describing the matter fields,
such that
\be
\E{M}{A} = \H{M}{\a} \Enull{\a}{A} \la{EHe}
\ee
where the intermediate indices split according to $\a= (\mu ,\ve)$.
For the explicit parametrization of the background frame,
we use the tangent space light-cone indices introduced above, viz.
$$
\Enull{\mu}{a} = \e{\mu}{a}  \;\;\; , \;\;\;
\Enull{\mu}{-} = u_\mu     \;\;\; , \;\;\;
\Enull{\mu}{+} = 0
$$
\be
\Enull{\ve}{a} = \Enull{\ve}{-} = 0   \;\;\; , \;\;\;
\Enull{\ve}{+} = 1   \la{Enull}
\ee
with inverse
$$
\Enull{a}{\mu} = \e{a}{\mu}  \;\;\; , \;\;\;
\Enull{-}{\mu} = n^\mu     \;\;\; , \;\;\;
\Enull{+}{\mu} = 0
$$
\be
\Enull{a}{\ve} = \Enull{-}{\ve} = 0   \;\;\; , \;\;\;
\Enull{+}{\ve} = 1   \la{Enullinv}
\ee
where $\e{\m}{a} n^\m = \e{a}{\m} u_\m = 0$ and
$n^\mu u_\mu =1$. The covariant and contravariant metrics
$g_{\m \n}$ and $h^{\m \n}$ in $d$ dimensions defined by
$$
g_{\mu \nu} = \Gnull_{\mu \nu} \equiv \Enull{\mu}{A}
\Enull{\nu}{B} \eta_{AB} =  \e{\mu}{a} \e{\nu}{b} \eta_{ab}
$$
\be
h^{\mu \nu} = \Gnull^{\mu \nu} \equiv \Enull{A}{\m} \Enull{B}{\n} \eta^{AB}
            =  \e{a}{\m} \e{b}{\n} \eta^{ab}  \la{gnull}
\ee
are therefore degenerate: $g_{\m \n} n^\n = h^{\m \n} u_\n =0$.
The projector onto the $(d-1)$-dimensional transverse subspace is
\be
{\Pi_\m}^\n := g_{\m \r} h^{\r \n} \equiv
  \e{\m}{a} \e{a}{\n}  \Longrightarrow     {\d_\m}^\n =
{\Pi_\m}^\n + u_\m n^\n   \la{transpro}
\ee
Geometrically the fibration by the orbits of the null
Killing field defines a projection from the $(d+1)$-dimensional
manifold onto the $d$-dimensional space of orbits. The usual Geroch
construction of tensors \ci{Ger} breaks down but one can still define
the image of the contravariant metric
by the projection map. As we said, it corresponds to
$h^{mn}$ and does not depend on a choice of section
{\sl i.e.} on the choice of the
coordinate $v$. Its determinant vanishes precisely when $\x^2$ does.
The choice of $n^m$ however is arbitrary and crucial to define
the transverse space to the fibration, the quasi-Maxwell field and the
covariant metric on the quotient space.

The ``matter" degrees of freedom are contained in
the $(d+1)$ by $(d+1)$ matrix $\H{M}{\a}$ with components
$$
\H{m}{\mu} = \d_m^{\;\; \mu} \;\;\; , \;\;\;  \H{m}{\ve} = S A_m
$$
\be
\H{v}{\mu} = 0 \;\;\; , \;\;\; \H{v}{\ve} = S   \la{H}
\ee
Consequently, the full vielbein is
$$
\E{m}{a} = \e{m}{a}  \;\;\; , \;\;\;
\E{m}{-} = u_m     \;\;\; , \;\;\;   \E{m}{+} = S A_m
$$
\be
\E{v}{a} = \E{v}{-} = 0   \;\;\; , \;\;\;
\E{v}{+} = S                   \la{E}
\ee
whose inverse we also record for completeness
$$
\E{a}{m} = \e{a}{m}  \;\;\; , \;\;\;
\E{-}{m} = n^m     \;\;\; , \;\;\;   \E{+}{m} = 0
$$
\be
\E{a}{v} = - \e{a}{m} A_m \;\;\; , \;\;\;
\E{-}{v} = - n^m A_m   \;\;\; , \;\;\;
\E{+}{v} = S^{-1}            \la{Einv}
\ee
We note that the background frame is recovered from $\E{M}{A}$ by
switching off the matter fields, i.e. by putting $A_m =0$ and
$S=1$ in these formulas. Then, of course,
$\xi_m \equiv u_m$ and $n^m \equiv N^m$, and there is no
need to distinguish intermediate from curved $d$-dimensional indices.
We shall nevertheless change frame by contracting tensors with the
appropriate frame matrix, keeping (usually) the name of the tensor as
is done traditionally in the Lorentz frame picture. A notable exception to
this rule will be $E$ itself.

Clearly we took the vector ${\bf E}_+$ along the Killing direction.
The vectors ${\bf E}_-$ and ${\bf E}_a$ complete the tangent vector
basis and ${\bf E}_-$ is to be chosen at will.
The full vielbein and its inverse are form invariant under the
subgroup $\ISO(d-1) \times \R$
of the Lorentz group. The $\R$ factor will be gauge fixed shortly
and reduced to a global subgroup. As we have mentioned the
very {\sl definition} of matter by the above
factorization is not $\ISO(d-1)$ invariant. This means that different
choices of the $n^m$ vector fields will lead to different
splittings between the matter field $A_m$ and the
background gravitational field.

The full $(d+1)$-metric has components
$$
G_{mn} = g_{mn} + S A_m u_n + S A_n u_m    \equiv
         g_{mn} + A_m \xi_n + A_n \xi_m
$$
\be
G_{mv} = S u_m \equiv \xi_m = W \p_mu \;\;\; , \;\;\; G_{vv} =0 \la{G}
\ee
where $\gnull_{mn} \equiv \e{m}{a} \e{n}{a} $.
Its inverse is
$$
G^{mn} \equiv h^{mn}  \;\;\; , \;\;\;
G^{mv} = S^{-1} n^m - h^{mn} A_n \equiv N^m
$$
\be
G^{vv} = h^{mn} A_m A_n - 2 S^{-1} n^m A_m \equiv N^v  \la{Ginv}
\ee

\subsection{Spacetime symmetries}
Let us first consider the symmetries preserving the choice of
``Lorentz" frames in the reduction from $(d+1)$ to $d$ dimensions; we
discuss them in some detail because of the new features that appear
in comparison with the usual non-null reduction.

Let us start with the
local Lorentz group $\SO (d,1)$, it is broken down to its subgroup
$\ISO (d-1) \times \R$ by the choice of gauge made in eq.\Ref{E};
this is the stability
subgroup of the flat $+$ direction or equivalently of the choice
 $\E{+}{m}=0$. It is to be
contrasted with the more familiar non-null reduction, where the
residual symmetry is $\SO (d)$ or $\SO (d-1,1)$. However, if we ignore
for the time being the $\R$ factor this
stability subgroup is isomorphic to the
Poincar\' e (Euclidean) group. The mathematical reason behind the
appearance of the Euclidean group here is related to the fact that the little
group of a null vector in Minkowskian geometry is the global Euclidean group.
As a local exact symmetry however, Euclidean invariance is rather unusual. In
ordinary general relativity, a local Poincare invariance is hidden which
can be
made explicit for example in total dimension three \ci{Wi} or
in four dimensions as the contraction of
a de Sitter gauge group \ci{Macdo}.
Our frame bundle has a priori a Lorentz structure group which can be reduced
to the $\ISO(d-1)$ subgroup in the presence of the Killing vector by our
choice of adapted frames. We do not
have to restrict it by some additional local assumption.

An important point to note is that the $(d+1)$-dimensional vector
representation of $\SO (d,1)$ is {\sl indecomposable} but not
irreducible under $\ISO (d-1)$ since it admits invariant
subspaces, but cannot be split. More explicitly, for an arbitrary
$\SO (d,1)$ covector $V_A = (V_a, V_-, V_+)$, we find that $V_+$ is
$\ISO (d-1)$ invariant, but that the variation of the components
$(V_a, V_-)$ contains terms involving $V_+$ and therefore they do not form
an invariant subspace under $\ISO (d-1)$.
To obtain a proper action of $\ISO (d-1)$ on this $d$-dimensional space,
we must quotient out the invariant subspace, or
equivalently impose the condition $V_+ =0$, which is
$\ISO (d-1)$ invariant and hence consistent.
Then the group $\ISO (d-1)$ acts on the $d$-dimensional space of covectors
$(V_a, V_-)$ and preserves the degenerate (contravariant)
metric $\eta^{ab} = \d^{ab},\;  \eta^{a-}= \eta^{--} =0$. Consequently it
preserves also $h^{mn}$, as well as $u_m:=\E{m}{-}$. In contrast neither
$n^m$ nor $\gnull_{mn}$ are preserved by the ``translation" generators of
$\ISO (d-1)$. The tensor calculus after setting to zero the $+$ component
would remain most analogous to the Lorentz
tensor calculus if we were to use only the contravariant metric.
(This would mean
in particular that the Lie algebra generators should have upper indices and
the parameters, connections and curvatures lower indices).

Let us now
consider the factor $\R$ corresponding to $(+-)$ boosts which preserve
the + direction as well.
If the action of the Lorentz generators is given by
$\d \E{M}{A} = \E{M}{B} L_B^{\;\; A}$, it is easy to see that
$\E{M}{-}$ and $\E{M}{+}$, i.e. $(u_m,0)$ and $(S A_m , S)$,
respectively, scale oppositely. This means that we could
boost the Kaluza Klein scalar $S$ away by setting
$S=1$. Instead we shall put $S=W$ in view of our previous result
\Ref{normality}, so that $u_m$ becomes the gradient of $u$.
This choice fixes the $\R$ factor of the Lorentz gauge subgroup,
after which we are left
with $\ISO(d-1)$ as the residual tangent space symmetry (times
the global scale invariance mentioned above).
Actually it will turn out that the {\sl boost rescaling} is not
the analog of the Weyl rescaling of dimensional reduction with
a non-null Killing vector. One may remark that the
Kaluza Klein scalar in the non-null case is inert under the
residual local Lorentz group (i.e. $\SO (d)$ or $\SO (d-1,1)$ for
\Ref{vielbein}). In the null case, it is
the residual local boost symmetry and the normality of the null Killing
vector established in the foregoing section which enable
us to find a Lorentz gauge where $u_m = \p_m u$ and which will thereby
permit the construction of torsion-free $\ISO (d-1)$ connections
in the following sections 4.3 and 5.1.

It is instructive to work out the action of $\ISO (d-1)$ on
all the components of \Ref{E}.
Denoting the $\ISO (d-1)$ parameters by
${L_a}^b$, ${L_a}^+$ and $L_-^{\;\; b} \equiv - {L_b}^+$, we have
$$
\d S =0  \;\; , \;\;  \d u_m =0
$$
\be
\d \e{m}{a} = \e{m}{b} {L_b}^a + u_m L_-^{\;\; a}
\;\; , \;\; \d A_m = S^{-1} \e{m}{a} {L_a}^+ \la{KK}
\ee
In other words we see that $A_m$ transforms
under the group  $\ISO(d-1)$, more precisely, it is contaminated by
the $d$-frame components.

Equation \Ref{KK} shows that by a further choice of Lorentz
gauge, we can achieve $A_a \equiv \e{a}{m} A_m = 0$, so that for this
particular choice of Lorentz frame that we call the ``anti-axial" gauge
$$
A_m = - \frc12 N^v  \xi_m
$$
and
\be
n^m=S N^m \la{covn}
\ee
The local group $\ISO (d-1)$ is thereby broken
to the transverse subgroup $\SO (d-1)$.

Let us note also that this is a convenient Lorentz
gauge for gravitational ``waves", or rather for Einstein solutions
admitting a null Killing vector; the explicit form of the
metric simplifies to\footnote{ If we further use $(d-1)$ transverse
but curved coordinates $x^i$ as well as $u$, we derive from $n^m=(n^i,1)$
that
$\gnull_{iu} = - \gnull_{ij} n^j$ and $\gnull_{uu} =n^i\gnull_{ij} n^j $.
In fact one could choose $n^i=0$
instead of $A^a=0$.}
\be
ds^2 = \gnull_{mn} dx^m dx^n - W^2 N^v du^2 + 2W du dv
\ee
The $pp$-wave metric corresponds to the case $W=1$ \ci{Kun}.

It is equally instructive to list the $\ISO(d-1)$ transformation rules of
the inverse frame:
$$
\d \e{a}{m}= -{L_a}^b \e{b}{m} \;\;, \;\; \d n^m= -L_-^{\;\; b} \e{b}{m}
$$
\be
\d S^{-1} =0\;\;,\;\; \d A_a = -{L_-}^a S^{-1}-{L_a}^b A_b\;\; ,
\;\; \d A_- = -{L_-}^b A_b \la{Kk}
\ee

These formulas show a splitting between what one could call matter, and
background gravitational fields. But this splitting depends
on the frame! The flat components of the quasi-Maxwell
gauge field that appear above have a ``covariantized" $\veps$-transformation
rule. The new geometry will be discussed shortly but first
we would like to reexpress the previous dependences on the
choice of frame as a dependence on the choice of $n^m$.
We start from \Ref{KK}: $\d A_m = -S^{-1} \e{m}{a} {L_-}^a$.
In accordance with the
invariance of $G_{mn}$ we have $\d \gnull_{mn} = -2S\d A_{(m} u_{n)}$,
together with \Ref{Kk}: $\d n^m= -L_-^{\;\; b} \e{b}{m}$.
Let us now adopt a $d$ dimensional point of view. The conditions
$\gnull_{mp}n^p = 0$, $n^m u_m=1$ (for fixed $u^m$) are preserved
by the local $d$-dimensional frame transformations of
$n^m$, $A_m$ and $\gnull_{mp}$ of the form:
$$
\d n^m = -h^{mn}\lambda _n
$$
\be
\d A_m = -S^{-1}\lambda _m \;\; , \;\;
   \d \gnull_{mn} =u_m\lambda _n +u_n\lambda _m \la{newsym}
\ee
if $\lambda _p n^p=0$ i.e. $ \lambda _m = -\gnull_{mp} \d n^p$.
These are the translations of $\ISO(d-1)$
when we set $\lambda _n=L_-^{\;\;b}\e{n}{b}$.

Finally general coordinate transformations in $d+1$ dimensions act like
$\d V_M = \p_M \veps^N V_N + \veps^N \p_N V_M$ on a covector $V_M$.
In accordance with general Kaluza Klein theory, one
would expect the original diffeomorphism invariance to reduce
to diffeomorphism invariance in $d$ dimensions times an ordinary
abelian gauge invariance of the vector field $A_m$. Indeed, it
is easy to check that the Maxwell-type gauge transformations
are identified with general coordinate transformations along the
$v$ direction, i.e. $\veps^M (x) = (0,...,0,\veps^v(x)\equiv \veps(x))$,
hence the name $\veps$-transformations. They read:
\be
x'^m (x,v)=x^m \ ,\ v'(x,v) =v - \veps (x^m) \ ,\
    A'_m(x) = A_m (x) + \partial_m \veps (x).\la{Max}
\ee
Given a frame the corresponding one-form
\be
E^+:=\E{M}{+} dx^M = S(dv+A_mdx^m) = \H{M}{\ve} dx^M \la{HMdxM}
\ee
is  by construction $\veps$-invariant,
it lives on the full (fibered) space
and is associated to a ``horizontal-vertical'' splitting. Consequently,
only the field strengths $A_{mn} \equiv \p_m A_n - \p_n A_m$
will appear in the equations of motion.
It can be checked that both $\xi_m$ and the contravariant
metric $G^{mn} \equiv h^{mn}$ are $\veps$-invariant, whereas
$G_{mn}$ in \Ref{G} is obviously not. Observe that
the difference between $G_{mn}$ and the
$\veps$-invariant (but degenerate) metric $\gnull_{mn}$ (cf.
\Ref{G}) involves two different vector fields. This is due to the fact that
the $\veps$-gauge field $A_m$ here is {\sl not} the same
as the Killing vector $\xi_m$ unlike in ordinary Kaluza Klein theory.
In fact this is just another way of saying that the covariant metric
does not project onto the orbit space.
To summarize, the main difference with non-null dimensional reduction
is that the quasi-Maxwell field depends on the frame, we shall have to
combine it with other fields in order to obtain Lorentz invariant objects.

As far as the intermediate frame is concerned, we note that a priori
the decomposition \Ref{EHe} is invariant under (local)
${\rm GL} (d+1)$ transformations acting on the lower index $\a$,
if the upper index $\a$ transforms with the contragredient matrix.
However, in order to preserve $\H{m}{\mu} = \d_{m}^{\;\; \mu}$,
the action of a $d$-dimensional diffeomorphism
$x^m \rightarrow {x'}^m = {x'}^m (x^n)$ on the vector index $m$
must be accompanied by the same transformation acting on the
intermediate index $\mu$. Hence,
diffeomorphisms in $d$ dimensions acting on $\E{m}{A}$ are coupled
to linear (compensating) transformations acting on
$\Enull{\mu}{A}$, and the indices $\m ,...$ will
be regarded as  world indices of the $d$-manifold.
On the other hand, from \Ref{HMdxM} it can be seen that
the index $\ve$ is inert under $\veps$-gauge transformations
in contrast to $v$, which is not; this is the principal difference
between intermediate tensors and tensors referred to the curved
indices $M,N,...$. It is the choice of $v$ coordinate
(the choice of section) which introduces the gauge arbitrariness,
it is partly avoided by switching to intermediate (or Lorentz) indices.

With hindsight we could now return to  the $\veps$-variations of
the (frame independent) metric components and rederive from them the
parametrizations \Ref{G} and \Ref{Ginv} in terms
of a gauge field, we have
$$
G'_{mn}=G_{mn}+\x_m\p_n\veps+\x_n\p_m \veps
$$
$$
N'^m=N^m-h^{mn} \p_n \veps
$$
\be
N'^v=N^v - 2N^m \p_m \veps +h^{mn} \p_m \veps \p_n \veps \la{SEC}
\ee
and the other variations vanish. Assuming the transformation law
\Ref{Max} we can find where to introduce $A_m$ terms so as
to obtain \Ref{G} and \Ref{Ginv}.

\subsection{$(d+1)$-connection and curvature}
In $(d+1)$-dimensional space we shall consider the canonical torsionfree
and metric preserving affine connection. It is gauge equivalent to the
Lorentz connection as can be seen by going to the vielbein frame. The
intermediate moving frame introduced above allows us to describe the
{\sl same} connection in yet another linear gauge. The anholonomy
will contribute to the formulas of E. Cartan
giving the torsion and curvature tensors. In $d$ dimensions we shall use
the corresponding subframes but a different connection to be constructed
from the $(d+1)$-dimensional one.

The full vielbein conservation equation for
\Ref{E} in $d+1$ dimensions is
\be
\p_M \E{N}{A} + \o_{M\;\; B}^{\;\;\; A} \E{N}{B} =
 P_{MN}^{\;\; Q} \E{Q}{A}      \la{dE2}
\ee
where $\o_{MAB}$ and $P_{MN}^{\;\; Q}$ are the unique expressions for
the torsion-free
connection computed from \Ref{E} and \Ref{G} in the usual way.
Equation \Ref{dE2} can be rewritten as an expression of the Lorentz
reductibility of the affine connection and its holonomy:
\be
\p_M \E{N}{A}-P_{MN}^{\;\; Q} \E{Q}{A}=
- \o_{M\;\; B}^{\;\;\; A} \E{N}{B}
\ee
The intermediate frame analog of \Ref{dE2} defines the Lorentz
invariant $\Gamma $:
\be
\p_M \H{N}{\g}+ \Gamma_{M \b}^{\;\; \g}\H{N}{\b} =
P_{MN}^{\;\; Q} \H{Q}{\g}      \la{dEP}
\ee
and hence
\be
\Gamma_{\a \b}^{\;\; \g} \Enull{\g}{A} :=
\p_\a \Enull{\b}{A} + \o_{\a\;\; B}^{\;\;A} \Enull{\b}{B} \la{hermann}
\ee
where $\Gamma$ is not symmetric in general.
This frame has anholonomy, in analogy with the vielbein anholonomy:
\be
\Omega_{MN}^{\;\; A} := 2\p_{[M} \E{N]}{\;\; A}
\ee
we have
\be
\T_{MN}^{\;\; \a} :=  -2\p_{[M } \H{N]}{\a}.
\ee

We shall distinguish the matter free case by using
the connections $\onull $ and $\Pnull $ instead of their
generalisations $\o $ and $P $ to the case with matter. Namely
$\onull := \o ({\mkreis{E}})$ and $\Pnull := P({\mkreis{E}})$.
The vanishing of the torsion tensor reads for example in
intermediate coordinates:
\be
0=T_{\a\b}^{\;\; \g}:=2\G_{[\a\b]}^{\;\;\; \g} -\T_{\a\b}^{\;\; \g}
\ee
so $$\T_{\a\b}^{\;\; \g} = 2\G_{[\a\b]}^{\;\;\; \g}.$$
The curvature tensor is given by:
\be
R_{MNAB} (\o )= \p_M \o_{NAB} - \p_N \o_{MAB} +
  {\o_{MA}}^C \o_{NCB} - {\o_{NA}}^C \o_{MCB}  \la{Riemann1}
\ee
or equivalently by
\be
{R_{MNP}}^Q (P )= \p_M P_{NP}^{\;\; Q} - \p_N P_{MP}^{\;\; Q} +
P_{MR}^{\;\; Q} P_{NP}^{\;\; R} - P_{NR}^{\;\; Q} P_{MP}^{\;\; R}
\la{Riemann2}
\ee
and we have the equality $R_{MNPQ} \big(\o (E)\big) =
R_{MNPQ} \big( P(E)\big)$. Switching to intermediate indices, we get
\be
\H{\a}{M} \H{\b}{N} R_{MNAB} = - \Theta_{\a \b}^{\;\; \g}
    \o_{\g AB} + {\widehat R}_{\a \b AB}  \la{Riemann3}
\ee
where
\be
{\widehat R}_{\a \b AB} := \p_\a \o_{\b AB} - \p_\b \o_{\a AB} +
{\o_{\a A}}^C \o_{\b CB} - {\o_{\b A}}^C \o_{\a CB}
\ee
differs from \Ref{Riemann1} by an extra anholonomy term.
The Riemann curvature tensor in intermediate frame is given by
\be
R_{\a \b \g}^{\;\;\;\;\;\d} := \p_\a \G_{\b \g}^{\;\; \d}
         - \p_\b \G_{\a \g}^{\;\; \d} +
  \G_{\a \veps}^{\;\; \d} \G_{\b \g}^{\;\; \veps}
- \G_{\b \veps}^{\;\; \d} \G_{\a \g}^{\;\; \veps}
- \Theta_{\a \b}^{\;\; \veps} \G_{\veps \g}^{\;\; \d}. \la{Rinter}
\ee
A useful expression for the Ricci tensor is
\be
R_{\a\g} = \p_\a \G_{\b \g}^{\;\; \b}
     -\p_\b \G_{\a \g}^{\;\; \b}
+ \G_{\d \a}^{\;\; \b} \G_{\b \g}^{\;\; \d}
- \G_{\b \d}^{\;\; \b} \G_{\a \g}^{\;\; \d}  \la{Ricci}
\ee

\section{Background geometry and dimensional reduction}

{}From \Ref{KK} we know that $\ISO(d-1)$ invariance is broken by the
choice
$A_m=0$. Nevertheless we shall first consider the case $A_m=s=0$,
where $s=\log S= \log W$ is defined up to a constant that could be
reintroduced easily. For $S=1$, we obviously have $\xi_\mu = u_\mu,\,
N^\mu = n^\mu$.
Our goal in this section is to study the resulting
``matter-free'' geometry in $d$ dimensions that results
in this special case, and to
demonstrate that the null Killing reduction leads to a generalization of
the Galilean covariant theories already studied in
\ci{Tr,Do, Ku}. As shown there, generally covariant Galilean
theories of gravity are distinguished from the more familiar
relativistic ones in several ways.

The first two new features are the degeneracy of the (contravariant) metric
and the existence of a closed one-form in its
kernel. This form is required for the
definition of ``absolute time''. The closure of
this form can actually be shown to follow from
its conservation by parallel transport with respect to
a torsionless connection which is assumed to be compatible with the metric.
If there is a single null eigenvector, the closure follows after proper
normalization as well.
But even if one assumes the absence of torsion,
the affine Levi-Civita connection and the ``spin" connection
are not uniquely determined by requiring the vielbein to be
covariantly constant. The arbitrariness is parametrized by
a choice of two-form and
this third new feature can be traced back to the degeneracy of the
metric in $d$ dimensions as we will explicitly show. Furthermore the
connections in our moving frame approach seem to depend on the choice of
frame. The last new feature we may briefly mention at this stage is that one
needs to impose a condition on the curvature tensor in order to reduce its
number of independent components to the usual one of the Riemannian
situation. However we shall not impose the other restrictions needed to
recover Galilei-Newton theory, as they do {\sl not} follow from dimensional
reduction.

One difference between our treatment and most previous ones is that
we shall always keep in mind the $(d+1)$-dimensional origin of
the theory. Nonetheless we shall begin in $d$ dimensions by
first presenting an ``intrinsic'' analysis that makes no
reference to $(d+1)$ dimensions, recovering
and extending previous results by the use of moving frames.
We will then sharpen the analysis from a $(d+1)$-dimensional point of
view and show that the ambiguities afflicting the theory in $d$
dimensions can be entirely eliminated in this way. In particular,
by using a well chosen frame and the existence of a
non-degenerate metric and its associated
Levi-Civita connection in $d+1$ dimensions, we are led to an
associated Galilean connection in $d$ dimensions. We shall postpone
until the next section the study of its frame dependence.

Furthermore, unlike the authors of \ci{Du}, who came closest to our
purpose by considering a one-dimensional extension of
the Galilean spacetime they wanted to study, we do not
assume the covariant constancy of the null vector nor the existence of
a higher dimensional structure group different from the Lorentz group.
In our treatment, the normality property of the Killing vector
$\xi_m = W \p_m u$ together
with the boost rescaling
of the Kaluza Klein scalar $S$ will ensure the torsion-free
condition for the natural $\ISO (d-1)$ $d$-dimensional connection.
(This boost is supposed to have
been effected {\it before} the consideration of the matter free sector,
that will occupy us in this section; the very definition of
the scalar $S$ requires this partial gauge fixing). The possibility
to restrict the Lorentz structure group to its Euclidean subgroup is
locally guaranteed, as we said, by the existence of the Killing vector.

The reader who does not want to see the unavoidability
of the Galilean connection to be arrived at step by step in the next
two sections may now jump to section 5.3, where a shortcut allows us to
extract it ``from the blue". He (or she) will thus
miss the beauty of the moving
frame method, and the possibility to consider fermions.

\subsection{Galilean geometry in $d$ dimensions: moving frames.}

As we noted before there is no need to distinguish curved and
intermediate indices if the intermediate frame $H$ is just the unit
matrix \Ref{EHe}; furthermore even for $H\neq 1$, the derivative operators
$\p_m$ and $\p_\m = \H{\m}{M} \p_M$ have identical action on
$v$-independent quantities. In other words the intermediate
subframe can be considered as holonomic in $d$ dimensions.
We will consistently use
intermediate indices
from now on so as to facilitate the comparison with the
case treated in the following section where matter will be included
and to have manifest $\veps$-invariance.

We shall consider the $d \times d$ submatrix
$(\Enull{\mu}{a} , \Enull{\mu}{-} ) \equiv (\e{\mu}{a} , u_\mu)$ of
\Ref{Enull} as a {\sl Galilean}
frame with respect to the new {\sl holonomic} frame in the
$d$-dimensional reduced geometry. By this we mean simply that
\be
\Enull{\m}{A}h^{\m\n}\Enull{\n}{B} = \eta^{AB}
\ee
for $A$ or $B= a,-$.
We can introduce $\ISO (d-1)$-valued spin connection coefficients
$\oti_{\mu\;\; b}^{\;\; a}$ and $\oti_{\mu \;\;-}^{\;\; a}$. We define
the corresponding affine connection $\Gamti_{\m \n}^{\;\;\s}$ by
requiring the covariant constancy of the $d$-bein, i.e.
\be
\p_\m \e{\n}{a} + \oti_{\m\;\; b}^{\;\; a} \e{\n}{b} +
\oti_{\m \;\; -}^{\;\; a} u_\n
   = {\Gamti_{\m \n}}^{\;\; \s} \e{\s}{a}      \la{de}
\ee
\be
\p_\m u_\n = {\Gamti_{\m \n}}^{\;\; \s} u_\s   \la{dxi}
\ee
These equations are the moving frame extension of the
admissibility conditions of a Galilean connection. Observe the absence of
spin connection terms
on the left hand side of \Ref{dxi}, it is due to our insistence
on $\ISO (d-1)$ as the proper tangent space group rather than
$\SO (d)$ or $\SO (d-1,1)$. It can be rewritten as
$D_\m (\Gamti ) u_\n =0$, while the other equation expresses the
conservation of the degenerate metric $h^{\m\n}$, i.e.
$D_\m (\Gamti ) h^{\nu \rho} =0$. We already alluded to the fact
that the Wigner-In\"on\"u contraction of $\SO (d)$ to
$\ISO(d-1)$ is easily implemented by first arranging the indices of all
the metric tensors to be in the upper position, and then by replacing
the unit metric by the once degenerate $d \times d$ submatrix of
$ \eta^{AB}$. Then our tensor calculus
is almost unchanged. Note that our requirements \Ref{de} and \Ref{dxi}
imply the conservation of the usual antisymmetric tensor densities of
order d.

The linear system of equations \Ref{de} and \Ref{dxi} can be solved in
the usual
fashion, apart from certain ambiguities which we will now exhibit.
Equation \Ref{dxi} implies that the torsion
$\Tti_{\m \n}^{\;\; \r} \equiv 2\Gamti_{[\m \n]}^{\;\; \r}$ obeys
\be
u_\r \Tti_{\m \n}^{\;\; \r} = u_{\m \n} \equiv \p_\m u_\n - \p_\n u_\m
\la{torsion}
\ee
A torsion-free geometry thus obtains if and only if $u_{\m \n} =0$.
We have already shown that this condition can always be satisfied by an
appropriate boost rescaling of the Kaluza Klein scalar $S$ if
the normality property \Ref{normality} of the Killing vector holds.
So our choice of zero torsion had two strong implications: it forced us
to assume one equation of motion so as to obtain \Ref{normality} and
then it was used to fix one boost generator of the residual Lorentz
gauge subgroup.

So let us proceed to the solution; multiplying \Ref{de} by
$\e{\r}{a}$ and symmetrizing in the indices $(\r \n)$ one finds:
\be
{\Gamti_{\m (\n}}^{\;\;\; \s} \gnull_{\r) \s} =
\frc12 \p_{\m} \gnull_{\n \r} - \oti_{\m - (\r} u_{\n)}.
\ee
This projection complements \Ref{dxi} and allows a complete computation
of $\Gamti $. From this relation we also see that $\gnull$ is {\sl not}
covariantly constant, unlike $h$. Instead, we have
\be
D(\Gamti)_\r \gnull_{\m\n} = 2 \oti_{\r-(\m} u_{\n)}. \la{DGg}
\ee

As far as $\oti $ is concerned the projection is faithful as it
does not have any $\oti_{m \; \n}^{\; -}$ component.
Taking cyclic permutations in the usual way, we get
\be
 {\Gamti_{\m \n}}^{\;\; \s} \gnull_{\s \r} =
-\frc12 \p_\r \gnull_{\m \n} + \p_{(\m} \gnull_{\n )\r}
         + \oti_{\r -(\m} u_{\n)}
 - \oti_{\m -(\n} u_{\r)} - \oti_{\n -(\r} u_{\m )} \la{Gammamnp}
\ee
Contracting with the contravariant metric $h^{\s \r}$, taking
into account \Ref{transpro}, \Ref{dxi} and renaming indices, we arrive at
the fundamental formula
\be
{\Gamti_{\m \n}}^{\;\; \r} =
{\Gamti_{\m \n}}^{\;\; \r} (n,K,h) := {\Gamti_{\m \n}}^{\;\; \r}(n)
   + 2 u_{(\m} K_{\n )\s} h^{\s \r}     \la{Gammamnp1}
\ee
where
\be
{\Gamti_{\m \n}}^{\;\; \r}(n) := \frc12 h^{\r \s} \Big(
 2 \p_{(\m} \gnull_{\n) \s} - \p_\s \gnull_{\m \n} \Big)
 + n^\r \p_{(\m} u_{\n )}\la{Pnull}
\ee
and
\be
K_{\m \n} := \oti_{[\m \n]-}. \la{Kmn}
\ee
${\Gamti_{\m \n}}^{\;\; \r}(n)$ is the analog of the usual torsion-free
affine connection (Christoffel symbol). We shall check later (see eq.
\Ref{PPP}, \Ref{Yeah})
that ${\Gamti_{\m \n}}^{\;\; \r}(n)$ is
equal to $\Pnull_{\m \n}^{\;\; \r}$,
the $d$-dimensional part of the $(d+1)$-dimensional Christoffel symbol.

Owing to the degeneracy of our system, the antisymmetric
tensor $K_{\m \n}$ defined by \Ref{Kmn}
can be chosen arbitrarily in \Ref{Gammamnp1}.
This means that the affine connection coefficients
are not uniquely determined by \Ref{de} and \Ref{dxi}. This is a typical
situation for a degenerate inhomogeneous linear system: when it admits
one solution, it admits many. The $\ISO(d-1)$ action on the vector field
$n^\m$ via \Ref{newsym} could be compensated in our $\Gamti (n,K,h)$
if we had (for $S=1$)
\be
\d K_{\m \n} = - \p_{[\m} \lambda _{\n]} \la{delKK}
\ee

In the same fashion one can solve \Ref{de} for the spin connection
coefficients, where similar ambiguities are encountered.
Contracting \Ref{Gammamnp} with $n^\r$ and using
$n^\n \gnull_{\m \n} =0$ we find
\be
\oti_{(\m \n)-} = \frc12 n^\r {\cL}_\r \gnull_{\m \n}
         + 2 n^\r K_{\r (\m} u_{\n )}
\la{omn-}        \ee
where
\be
n^\r {\cL}_\r \gnull_{\m \n} := n^\r \p_\r \gnull_{\m \n} +
          2(\p_{(\m} n^\r) \gnull_{\n )\r}    \la{Lieder}
\ee
is the Lie derivative.
The coefficients of anholonomy read:
\be
\Oti _{ab}^{\;\; c} =: 2 \e{a}{\m} \e{b}{\n} \p_{[\m} \e{\n ]}{c}   \;\, , \;\;
\Oti _{-b}^{\;\; c} =: 2 n^\m \e{b}{\n} \p_{[\m} \e{\n ]}{c}   \la{Onull}
\ee
(the other coefficients involve $\p_{[\m} u_{\n]}$  and vanish in the
background if the torsion is set equal to zero),
we get the familiar formula for the transverse components
\be
\oti_{abc} = \frc12 \Big( \Oti_{abc}
   - \Oti _{bca} + \Oti _{cab} \Big)   \la{onullabc}
\ee
The remaining components of the
spin connection in flat indices are given for arbitrary $K_{\m\n}$ by
$$
\oti_{[ab]-} = K_{ab} = \frc12 \Oti _{ab-} \;\; , \;\;
\oti_{-a-} = 2 K_{-a}
$$
\be
\oti_{(bc)-} = \Oti _{-(bc)} \;\; , \;\;
\oti_{-bc} = \Oti _{-[bc]} - K_{bc}  \la{onullbc-}
\ee
where indices have been converted from flat to curved
by means of the $d$-bein
$\Enull{\m}{a,-}$; so $K_{ab} \equiv \e{a}{\mu} \e{b}{\nu} K_{\mu \nu}$
 and $ K_{-a} \equiv n^\mu \e{a}{\nu} K_{\mu \nu}$ .
These equations are, of course, consistent with \Ref{Kmn} and
\Ref{omn-}.

In summary, the torsion-free parallel transport conditions for the
background $d$-frame are solvable provided $u_{[\m\n]}=0$, and involve an
arbitrary antisymmetric tensor $K_{\m\n}$ or equivalently an arbitrary
choice of $\oti_{[\m \n]-} $ or equivalently an arbitrary choice of
the $-$ component of the $\ISO(d-1)$ connection one-form $\oti_\m$.

\subsection{Previous work on Galilean geometry}

We are now ready to establish the connection with previous work
on the differential geometry of Galilean covariant theories when
$A_m =0$ and $S=1$; the general case with {\sl matter}
will be treated in the following section.
It appears (see \ci{Eh2}) that the geometric structure
emerged from the work of \ci{We}. In \ci{Do}
the choice of a ``field of observers'' (our $n^m$)
was shown to be related to the determination
of a covariant metric tensor. These authors carefully discussed
Newton's laws and the so-called {\sl special} connections associated
to the various (fields of) observers
whose worldlines are tangent to the $n^\m$ vector fields.
In fact these special connections are simply our \Ref{Pnull}. They
can be interpreted as incorporating not only potential but also
Coriolis (or a subset of Lorentz-type) forces. In other words, these
forces can be hidden by a suitable change of observers.
This is a generalized Galilean equivalence principle.

The special connection $\Gamti (n)$ admits $n$ as
a geodesic affinely parametrized vector field:
\be
n^\m D(\Gamti (n))_\m n^\r = 0 .   \la{nDn}
\ee
It is in fact characterized by this property, and the constraint
\be
D(\Gamti (n))^{[\m} n^{\r]} = 0.  \la{Dn}
\ee
where the index has been raised with the degenerate metric $h$ \ci{K2}.
What was
not clear in previous works was the reason for the identification of the
Lorentz force with inertial effects,
i.e. changes of observers. It will appear here as a consequence of
the Lorentz invariance of the original theory and the existence of
one more null direction.

The most general connections that preserve the space foliation of
spacetime  with its metric on the leaves are called {\sl admissible},
they are our \Ref{Gammamnp1} with {\sl arbitrary} $K_{\m \n}$.
In \ci{Ku} these results were combined with other physicists' work,
see in particular \ci{Tr}. The ``Galilean'' structure that emerged
can be characterized by a degenerate contravariant metric
$h^{\m\n}$, a foliation with normal $u_\m$ which is a closed
one-form in the kernel of $h^{\m\n}$ and the set of torsion-free affine
connections preserving the metric and 1-form $u_\m dx^\m$: the
admissible connections.

These data correspond to ours: the contravariant Galilean metric
is the same as our $h^{\m\n}$, and the vector defining the foliation
is our $\Enull{\m}{-}$, it is proportional to $\xi_\m$, which is
indeed in the kernel of $h^{\m\n}$.
As we already pointed out, the absence of torsion and the Galilean
structure had to be imposed
by hand in previous works, whereas no such assumption needs to be made
if one starts from a higher dimension. Then the existence of an
absolute time and the absence
of torsion follow from the Kaluza Klein reduction by requiring
one of Einstein's equations of motion. The special connection
associated to the field $n^m$ is nothing but \Ref{Pnull}, and in fact
the correspondence goes further.

In \ci{Ku} the ambiguity in the structure preserving affine connections,
i.e. the difference between two admissible connections is
explicitly parametrized
by a two-form $K_{\m \n}$ as we also showed in \Ref{Kmn}. But
in \ci{Tr, Ku}, it was noticed that an extra restriction is
needed if the curvature tensor of the Galilean theory is to have
the same number of independent components as the usual relativistic
one (since otherwise, the Galilean theory could not correspond to
the $c\rightarrow \infty$ limit of a matter free relativistic one). This
led to what we shall call the ``Newton Coriolis" (NC-)condition:
\be
\Rti_{\m \;\; \n \;\;}^{\;\; \r \;\; \s} (\Gamti, h) =
\Rti_{\n \;\; \m \;\;}^{\;\; \s \;\; \r} (\Gamti, h)  \la{Newton}
\ee
Recall that in the Minkowskian case, this relation follows from the
torsion Bianchi identity; here it imposes non-trivial restrictions
because the contravariant metric is degenerate. We may use the identity
\be
\Rti_{[\m \;\; \n] \;\;}^{\;\; (\r \;\; \s)} \equiv 0.
\ee
so the condition \Ref{Newton} is sometimes written
\be
\Rti_{(\m \;\; \n) \;\;}^{\;\; [\r \;\; \s]} =0.
\ee
As shown in \ci{Ku},
the condition \Ref{Newton} is satisfied by ``special" connections.
So among the ``admissible" connections
${\Gamti_{\m \n}}^{\;\; \r} (n,K,h)$ (see eq.\Ref{Gammamnp1}),
the NC-connections are those for which $\p_{[\m} K_{\n\r]} = 0$.
This makes sense because
changing $n^\m$ to $n'^\m $ corresponds to changing:
\be
{\Gamti_{\m \n}}^{\;\; \r}(n')={\Gamti_{\m \n}}^{\;\; \r}(n)
+2 u_{(\m} K_{\n )\s}(n,n') h^{\s \r}. \la{Cov1}
\ee
where the two-form
\be
K_{\n\s} (n,n')=\p_{[\n}\lambda _{\s]} \la{Cov2}
\ee
is obviously closed ($\lambda _\m$ was introduced in \Ref{newsym}, it
satisfies $\lambda .n = 0$ and we have
$n'^\r -  n^\r = -h^{\r\m} \lambda _\m $).
The closed two-form $K$ can be shown to
produce a Lorentz-type force in the equations of motion
for a point particle. It can be
reabsorbed by a suitable change of frame $\d n$ if the vector potential
is normal to $n$, namely if $K$ is of the form \Ref{Cov2}.

However, to recover the true Newtonian limit corresponding to a
potential force and to ensure the absence of Coriolis-type
forces, \Ref{Newton} is {\sl not enough}. Rather, one must impose
besides the automatic volume preservation:
\be
\Rti_{\r\s\t}^{\;\;\;\;\;\;\t} =0
\ee
further conditions on the connections \ci{Tr, Tr2, Di, Eh2},
for example a dynamical one:
\be
\Rti_{\r\s[\n}^{\;\;\;\;\;\;\t} u_{\m]} = 0
\ee
It corresponds to the existence of $(d-1)$ covariantly constant vector
fields tangent to the spacelike slices. It implies the Galilean analog of
Einstein's equations:
\be
\Rti_{\m \n} = \rho u_\m u_\n.     \la{GalEinst}
\ee
One may also assume other dynamical,
as opposed to kinematical, constraints. Typically the Galilean analog of
Einstein's equations is supposed to have the form \Ref{GalEinst}
with $\rho$ the mass density. We will see in section 6 that the
equations of motion obtained by dimensional reduction from
$d+1$ dimensions are {\sl not} of this form if the dilaton field
is included. Equation \Ref{GalEinst} implies in particular that the
equal time sections are Ricci flat, and hence that for $d \leq 3$ one can
choose flat ``Galilean" coordinates. In this case, Einstein's
equations can be rewritten as a non-linear modification of
Maxwell's equations in flat
space \ci{Eh2}. For $d>3$, on
the other hand, the vanishing of the Ricci tensor no longer implies
that the full Riemann tensor is zero, so the space manifold need
not be flat, and there may be genuine gravitational effects
in addition. The problem of
identifying a gravitational action for Galilean gravity
will be discussed in section 6.2.


\subsection{$\ISO (d-1)$ connections from $d+1$ dimensions}
We shall now reconsider the results of subsection $4.1$ and explain
how the formulas \Ref{de} and \Ref{dxi}
and their solutions \Ref{Gammamnp1}, \Ref{onullabc} and \Ref{onullbc-}
can be re-interpreted from a $(d+1)$-dimensional point of view.
Of course, the flat and curved metrics in $(d+1)$ dimensions
are no longer degenerate, and thus the situation is unambiguous. As
there is more structure at hand we might
expect to simply exhibit a particular $d$-connection.
In fact, through this procedure, we can
determine canonically the two-form $K_{\m \n}$ corresponding to a choice
of $n^m$.

Let us therefore consider the analog of \Ref{de} and \Ref{dxi}
in $d+1$ dimensions. We keep the same notations but the tangent space
group is now $\SO (d,1)$ for $\onull$.
\Ref{hermann} becomes
\be
\p_\a \Enull{\b}{A} + \onull_{\a\;\; B}^{\;\;A} \Enull{\b}{B} =
\Pnull_{\a \b}^{\;\; \g} \Enull{\g}{A} \la{dEnull0}
\ee
where the superscript $(\circ)$ indicates that the corresponding
quantities are given by the standard expressions computed from the
background vielbein $\Enull{\a}{A}$ in the usual way; in particular,
$\Pnull_{\a \b}^{\;\; \g}$ is the unique torsion-free affine connection
in $d+1$ dimensions that preserves the metric $\Gnull_{\a \b}$.
We next substitute the explicit
form of \Ref{Enull} and write out the components corresponding
to $A=(a,-,+)$ explicitly. In this way, we obtain three equations, viz.
\be
\p_\a  \Enull{\b}{a} + \onull_{\a\;\; b}^{\;\; a} \Enull{\b}{b}
         + \onull_{\a\;\; -}^{\;\; a}  \Enull{\b}{-}
         + \onull_{\a\;\; +}^{\;\; a}  \Enull{\b}{+}=
\Pnull_{\a \b}^{\;\; \m} \Enull{\m}{a}   \la{dEnull1}
\ee
\be
\p_\a  \Enull{\b}{-} + \onull_{\a\;\; b}^{\;\; -} \Enull{\b}{b}
         + \onull_{\a\;\; -}^{\;\; -}  \Enull{\b}{-} =
\Pnull_{\a \b}^{\;\; \m} \Enull{\m}{-}   \la{dEnull2}
\ee
\be
\p_\a  \Enull{\b}{+} + \onull_{\a\;\; b}^{\;\; +} \Enull{\b}{b}
         + \onull_{\a \;\; +}^{\;\; +}  \Enull{\b}{+} =
\Pnull_{\a \b}^{\;\; \ve} \Enull{\ve}{+}   \la{dEnull3}
\ee
where, on the right hand side, we took into account that
$$\Enull{\m}{+} = \Enull{\ve}{a} = \Enull{\ve}{-} =0$$
 (cf. \Ref{Enull}).
We should keep in mind that, on the left hand
side, only the derivatives $\p_\a$ with $\a = \m$ contribute
because $\p_v \equiv \p_\ve \equiv 0$ by dimensional reduction.

We would like now to compute $\Gamti_{\m\n}^{\;\; \r}$. Let us compare
\be
\Gamnull_{\a \b}^{\;\; \g} \equiv
\Pnull_{\a \b}^{\;\; \g} \equiv \frc12 \Gnull^{\g \d}
\big( 2 \p_{(\a} \Gnull_{\b )\d} - \p_\d \Gnull_{\a \b} \big)  \la{PPP}
\ee
to
\Ref{Gammamnp1} for $(\a,\b,\g) = (\m,\n,\r)$. We have the
identity
\be
\Pnull_{\m\n}^{\;\; \r} = \Gamti_{\m\n}^{\;\; \r} (n). \la{Yeah}
\ee
Comparing \Ref{de} with \Ref{dEnull1},
we immediately see that they are compatible with
\be
\Gamti_{\m\n}^{\;\; \r}=\Pnull_{\m\n}^{\;\; \r} \la{Yep}
\ee
i.e.  $K_{\m \n} =0$ in \Ref{Gammamnp1}.

Equation \Ref{dEnull2}, on the other hand,
differs from \Ref{dxi} by extra spin connection terms,
which are not $\ISO (d-1)$ valued. To relate \Ref{dxi} to
the corresponding components of
\Ref{dEnull2}, {\it the trick is to shift} the extra terms from the
left to the right hand side, in such a way that \Ref{dEnull2} becomes
\be
\p_\a \Enull{\b}{-} =
\Pnull_{\a \b}^{\;\; \mu} \Enull{\mu}{-}  -
   \onull_{\a \;\; A}^{\;\; -}  \Enull{\b}{A}, \la{dEnull4}
\ee
and re-interpret the right hand side as a new affine connection.
Since the resulting expression is no longer manifestly symmetric in
$(\a ,\b )$ when $u_{\m \n} \neq 0$, the emergence of
$d$-dimensional torsion is a possible dangerous consequence of
this rearrangement. It is this choice not to introduce torsion that leads
us at this stage to use one of the original equations of motion to enforce
the condition $u_{\m \n}=0$ as anticipated in section 3.2.
 In the case at hand however, the spin connection components that are not
$\ISO (d-1)$ valued
($\onull_{\a \;\; A}^{\;\; -} = \onull_{\a+A}$, of which we need
only $\onull_{\m+A}$) are simply absent when $u_{\m \n} =0$.
And we have
\be
\p_\a \Enull{\b}{-} = \Pnull_{\a \b}^{\;\; \mu} \Enull{\mu}{-}. \la{minus}
\ee

Finally \Ref{Kmn} can be rewritten
$K_{\m \n} := \frc12 \Onull_{\m \n}^+=0$,
and hence $K$ vanishes in the absence of ``matter".
\Ref{dEnull3} has no counterpart in $d$ dimensions, and the same remark
applies to the other components of \Ref{dEnull1} and \Ref{dEnull2}.

In summary, starting from the unique $(d+1)$ connection and an explicit
choice of frame, we
obtained $K_{\m \n} =0$ and thereby the simplest possible
$d$-dimensional NC-connection \Ref{Yep}.
In the following chapter, we will extend these considerations
to the case where Kaluza Klein matter is included and again obtain
a canonical NC-connection.

\section{Kaluza Klein matter couplings}

We will now switch on the {\sl matter} fields residing in the
intermediate frame $\H{M}{\a}$, whose inverse we denote by $\H{\a}{M}$.
We will use the intermediate frame and its inverse to convert
world indices into intermediate ones and vice versa. Their use will
considerably simplify the computations by comparison to the use of the
``Lorentz" frame.
By construction (see remarks at
the end of section 3), the intermediate frame equations to be written
below are still manifestly $\veps$-gauge invariant and turn out
to be compatible
with the vanishing of torsion in $d$ dimensions. By descending
from $d+1$ dimensions, a unique $d$ connection can be constructed,
and it is really part of the $(d+1)$ connection in disguise.

We follow a strategy that will permit us to directly compare
the connections that are obtained in presence of matter with the
background connections in \Ref{dEnull0} and \Ref{Yep}, and to read off
the two-form $K_{\m \n}$ from the equations. We recall the equations
\Ref{hermann} for $\Enull{\a}{A}$,
where now  the spin connection $\o_{\a AB}$ and the affine
connection $\Gamma_{\alpha \beta}^{\;\;\gamma}$ differ from the corresponding
expressions for the pure background by extra terms depending on the
Kaluza Klein matter fields $S$ and $A_\m$. (The latter field was
actually considered in \ci{Du} implicitly). Here we are in a pure
Kaluza Klein situation and we shall not make any other assumption than the
existence of the null Killing vector, admittedly at the cost of some
complications.
In ordinary Kaluza Klein dimensional reduction one witnesses the emergence
of a scalar field and a vector gauge field in $d$ dimensions. But here,
although it was known that there are several Galilean approximations
to Maxwellian electromagnetism \ci{LLB}, one finds a {\sl nonlinear} version
of electromagnetism that will be in
some sense {\sl hidden} inside a generalized gravitation theory, in
this connection see also \ci{Eh2, Bel}.

\subsection{Differential geometry with intermediate frames}
Let us now apply the results of section 3 for the various frames and
connections, more precisely the various component descriptions
of the canonical Riemannian connection. The covariance of the full
vielbein
\Ref{E} in $d+1$ dimensions is expressed by equation \Ref{dE2},
where $\o_{MAB}$ and $P_{MN}^{\;\; Q}$ are the complete expressions for
the (torsion-free)
connection computed from \Ref{E} and \Ref{G} in the usual way.
We now substitute \Ref{EHe} into \Ref{dE2} and move the
terms with derivatives on $\H{M}{\a}$ to the right hand side.
The result is a gauge transformed version of \Ref{hermann}:
\be
\Gamma_{\a \b}^{\;\; \g} = P_{\a \b}^{\;\; \g} +
      \p_\a \H{\b}{N} \H{N}{\g}   \la{GisPplusdH}
\ee
where, of course, $P_{\a \b}^{\;\; \g} \equiv \H{\a}{M} \H{\b}{N}
P_{MN}^{\;\; Q} \H{Q}{\g}$.
It is straightforward to check that $\p_\a \H{\b}{N} \H{N}{\g} = 0$
unless $\g = \ve$. The absence of torsion implies
\be
\T_{\a \b}^{\;\; \g} := 2\G_{[ \a \b ]}^{\;\; \g}
       = 2\p_{[\a } \H{\b]}{N} \H{N}{\g}.  \la{Theta}
\ee

The non-vanishing components of the anholonomy associated with
the intermediate frame $\H{M}{\a}$ are found to be
\be
\T_{\mu \nu}^{\;\; \ve} = -S A_{\mu \nu} \;\; , \;\;
\T_{\mu \ve}^{\;\; \ve} = S \p_\mu S^{-1}  \la{Theta1}
\ee
This shows that the $\a\b$ indices of  $\G_{\a \b}^{\;\; \ve}$
do not appear symmetrically (although the torsion tensor
is still zero) when the matter fields are switched on, whereas
the purely $d$-dimensional components $\G_{\m \n}^{\;\; \r}$
remain symmetrical. Pulling down the index with the
background metric ($\Gnull_{\a\b}\equiv \G_{\a\b}$), we get
\be
\T_{\m \n , \r} =  - A_{\m \n} \xi_\r \;\; , \;\;
\T_{\m \ve , \r} =  \p_\m S^{-1} \xi_\r   \la{Theta2}
\ee
and
\be
\T_{\m \n ,\ve} = \T_{\m \ve ,\ve} =0.
\ee
These formulas can be combined into a covariant equation:
\be
\T_{\a \b ,\gamma} = - \xi_\gamma A_{\a \b}
\ee
provided we define $A_\ve = -S^{-1}$ as one may infer from \Ref{Einv}.

Next, we determine the symmetric part of the affine connection.
It is natural to define
\be
2 \th_{\a \b \g} := \Theta_{\a \b \g} - \Theta_{\b \g \a} +
                 \Theta_{\g \a \b}    \la{S}
\ee
(compare to eq.\Ref{onullabc}) and notice that
\be
\G_{\a \b }^{\;\; \g} = \Pnull_{\a \b}^{\;\; \g}+
\th_{\a \b }^{\;\; \g} \la{Jul}
\ee
because we have \Ref{PPP}, \Ref{Theta} together with its symmetric partner:
\be
2\G_{M(\a\b)} = \p_M\Gnull_{\a\b}.
\ee
$\th$ and $\G$ are both equal to the usual Lorentz connection when the
background frame is trivial. In general, $\th$ and $\Pnull$,
respectively, may be characterized as the terms of $\G$ that
contain derivatives of $\H{M}{\a}$ and $\Gnull$, respectively.
In terms of $\T$ we obtain:
\be
\G_{(\a \b )}^{\;\; \g} = \Pnull_{\a \b}^{\;\; \g} -
\Gnull_{\d (\a} \Theta_{\b) \veps}^{\;\; \d} \Gnull^{\veps \g}
\la{Gsymm}
\ee
and, together with \Ref{Theta},
\be
\G_{\a \b}^{\;\; \g} = \Pnull_{\a \b}^{\;\; \g} -
\Gnull_{\d (\a} \Theta_{\b) \veps}^{\;\; \d} \Gnull^{\veps \g}
+ \frc12 \Theta_{\a \b}^{\;\; \g}      \la{fullG}
\ee
These expressions are $\veps$-gauge invariant as anticipated.

We can also solve \Ref{hermann} for the spin connection. The full
spin connection is given by
\be
\o_{ABC} = \frc12 \Big( \O_{ABC} - \O_{BCA} + \O_{CAB} \Big)
\ee
where
\be
\O_{ABC} = \Onull_{ABC} +  \Theta_{ABC}
\ee
and, of course, ${\Theta_{AB}}^C = \Enull{A}{\a} \Enull{B}{\b}
\Theta_{\a \b}^{\;\; \g} \Enull{\g}{C}$. Using \Ref{S}
we find
\be
\o_{ABC} = \onull_{ABC} + \th_{ABC}. \la{onullplusS}
\ee

We keep the same notation $\Gamti$ for the $d$-dimensional connection
we are looking for.
Let us again compare $\G_{\mu \nu}^{\;\; \rho}$, with its
indices restricted to $d$ dimensions, to
${\Gamti_{\m \n}}^{\;\; \r}(n,K)$ which we have seen in \Ref{Gammamnp1}
to be the most general connection compatible with our degenerate
geometry. From \Ref{fullG}, we find for the corresponding
components of the $(d+1)$-dimensional connection
\be
\G_{\mu \nu }^{\;\; \rho} = \Pnull_{\mu \nu }^{\;\; \rho}
  +n^\rho u_{(\m} \p_{\n )} s +
S u_{(\mu} A_{\nu )\s} h^{\s \rho}   \la{Pmunurho}
\ee
Next we must define a $d$-dimensional $\ISO(d-1)$
connection and shift the unwanted components of the
Lorentz connection to the affine connection as
explained in subsection 4.3. Comparing \Ref{hermann} to
\Ref{de} and \Ref{dxi} we obtain tentatively in the same way as in
section 4.3 (see \Ref{Yep})
$$
\Gamti_{\m \n}^{\;\; a} = \G_{\m \n}^{\;\; a}
$$
\be
\Gamti_{\m \n}^{\;\; -} = \G_{\m \n}^{\;\; -} -
\o_{\m\;\; \n}^{\;\;\; -} =\p_{(\m} \Enull{\n)}{-}.
\ee
Hence
\be
\Gamti_{\mu \nu }^{\;\; -} = \Pnull_{\mu \nu }^{\;\; -}
\ee
where we used \Ref{minus}

Comparison with the $\rho \equiv -$ component of \Ref{Pmunurho} shows
that the second term on its right hand side drops out. Using
\Ref{Yeah} we find that the
$d$-connection is given by $\Gamti(n,K)$ of
eq.\Ref{Gammamnp1} with
\be
K_{\mu \nu} = \frc12 S A_{\mu \nu}.
\ee
Altogether, we are led to
\be
\Gamti_{\m \n}^{\;\; \r} =\Pnull_{\mu \nu }^{\;\; \rho}
   + u_{(\mu} SA_{\nu )\s} h^{\s \rho}=\Gamti(n,K,h). \la{GFINAL1}
\ee
Since $A_{\mu \nu}$ obeys the (Maxwell) Bianchi identity,
the two-form $K_{\mu \nu}$ is closed if $S$ is constant.

Remark: in this case the $d$-connection components are the same as the
$(d+1)$-dimensional ones:
\be
\Gamti_{mn}^{\;\; r} = \Gamma_{\m\n}^{\;\; \r} = P_{\m\n}^{\;\; \r}
= P_{mn}^{\;\; r}
\ee

On the other hand,
it appears at this point that the connection $\Gamti$ does {\sl not} satisfy
the NC property \Ref{Newton} any more if a non-constant scalar field is
included; moreover \Ref{GFINAL1} will be seen not to be invariant
under Lorentz transformations. In the next section we
will show how to cure both of these problems by taking into
account a Weyl-type rescaling while preserving $\veps$-gauge invariance.

\subsection{Weyl rescaling and $\ISO(d-1)$ gauge invariant connection}

{}From \Ref{GFINAL1} and \Ref{newsym} one can check that $\Gamti$ would
be invariant under the group $\ISO (d-1)$ if it were not for extra terms
involving derivatives of $s$. Recalling the variations
$\d n^\m$, $\d g_{\m \n}$, $\d \Gamti(n)$ and $\d A_\m$, we find
that all terms cancel except
$$
\d \Gamti_{\m \n}^{\;\; \r} = 2u_{(\m} \tilde \d K_{\n)\s}h^{\s\r}
$$
with
\be
\tilde \d K_{\n\s} :=\p_{[\n} v_{\s]}-S\p_{[\n} (S^{-1} v_{\s]})=
(\p_{[\n}s)  v_{\s]}. \la{mismatch}
\ee
Clearly the variation of our candidate Galilean connection
under $\ISO (d-1)$ transformations vanishes if and only if
$S$ is constant.

We now have two indications that our new field $S$ has introduced some
complications. This is to be contrasted with the case $S=1$ which was solved
with a natural NC-connection (in this special case one
recovers the result of \ci{Du}). From experience with ordinary
Kaluza Klein theories we know that problems with scalars usually
arise if the Weyl rescaling has not been properly taken into account.
In the case at hand, we notice that there is still one part of the
metric that remains at our disposal: the spatial part
($u_\m$ and $n^\m$ cannot be rescaled, because this would
reintroduce non-zero torsion). So let us define
\be
\gnull_{\m \n} =w g'_{\m \n}  \;\;\;  , \;\;\;
h^{\m \n}  = w^{-1} h'^{\m \n} \la{Weyl}
\ee
Inspection of \Ref{GFINAL1} now suggests that we should take
$S h^{\m \n} = h'^{\m \n}$, i.e. $w=W=S$. If we introduce a
Weyl rescaled parameter $\lambda'_\m := S^{-1} \lambda_\m$, we can
mimick the $S=1$ situation with the new
connection (see \Ref{Gammamnp1})
\be
\Pti := \Gamti \Big( n, K' =\frc12 A_{\m\n}, h' \Big)   \la{Ptilde}
\ee
that replaces \Ref{GFINAL1}. We could also have used the notation
$\Gamti'$ for the new connection but it is symmetrical and we chose to
call it $\Pti$.

Indeed provided we use the following Weyl rescaled version of \Ref{newsym}
that reads now:
\be
\d n^\m = - h'^{\m \n} \lambda'_\n  \;\; , \;\;
\d A_\m = - \lambda'_\m  \;\; , \;\;
\d g'_{\m \n} = u_\m \lambda'_\n + u_\n \lambda'_\m  \la{newsym1}
\ee
\Ref{Ptilde} is then invariant as
one can easily verify. However, the Weyl rescaling of the spatial
metrics gives rise to extra terms involving derivatives of $s$.
These do not affect the transformation properties of the connection
since they transform properly as tensors. Altogether we obtain for the
connection:
\be
2\Pti_{\m \n}^{\;\; \r} \equiv 2\G_{\m \n}^{\;\; \r} +
g'_{\m\n} h'^{\r\s} \p_\s s - 2 \d^\r_{(\m} \p_{\n )} s  ;
\la{FINALS}
\ee
where $\G_{\m \n}^{\;\, \r}$ are components of the $d+1$-connection.
This new connection preserves the Weyl rescaled contravariant metric
$h'^{\m \n} $ as well as $u_\m$, as it should be. We have emphasized
that the connection follows from \Ref{Gammamnp1} but with metric $h'$.
Its $\r \equiv -$ component differs from the previous tentative
connection $\Gamti$ by two more terms linear in $\p s$, so as to be
Lorentz invariant. $\Pti_{\m \n}^{\;\; \r}$ as given by \Ref{FINALS}
will be our final Lorentz invariant and $\veps$-gauge invariant connection,
which now satisfies also the NC-property \Ref{Newton} automatically,
as one can easily check.

For completeness, let us generalize an identity found for $S=1$ in
\ci{Du}, albeit stated in a somewhat different form there:
\be
\Gamti \Big( n^m, \frc12 A_{\m\n}, h' \Big) =
\Gamti \Big( N'^m , \frc12 u_{[m} \p_{n]} \Nti^v , h'\Big)
\ee
where we defined: $N'^M= S N^M$.
We may now interpret it as follows: the Lorentz invariance is completely
fixed by imposing the ``anti-axial" gauge $A_m \propto u_m$ of \Ref{covn}.
Then, one may think of the quasi-Maxwell field as containing {\sl Goldstone
fields} associated with the $(d-1)$ translation generators of the
$\ISO (d-1)$ subgroup of the Lorentz group. It will be interesting
to see the relevance of this observation for the possible
existence of hidden symmetries in the theory. Note that $N'^m$ is geodesic
if $N'^v$ is constant.

\subsection{Tensor calculus}

The question which has remained open until now
is whether there might not be
another way to identify the putative Maxwell degrees of freedom
in a completely covariant fashion. After all, our
attempts so far were based on a strategy which started from manifestly
$\veps$-covariant quantities, and then restored $\ISO (d-1)$
invariance step by step. Alternatively, let us now start from
manifestly $\ISO (d-1)$ invariant quantities and try to restore
$\veps$-covariance as well. The only such quantities depending explicitly
on the field $A_m$ are the components $G_{mn}$, $G^{vv} \equiv N^v$ and
$G^{mv} \equiv N^m $ of the inverse metric (cf. \Ref{metric},
\Ref{invmetric}, \Ref{G} and \Ref{Ginv}). We must now construct tensors for
this new invariance, the $\veps$-symmetry. Our
first attempt will be to try and rediscover the connection, and as
we will
adopt a $d$-dimensional point of view we shall use roman indices.

Let us assume we have found an $\veps$-covariant affine torsionless
connection $C$ that preserves the one-form $u_m$. The new symbol is temporary
as we do not know the connection a priori and do not use the previous
derivation. It seems natural
to begin with the covariantized analog of the Levi-Civita connection:
\be
L_{mn}^{\;\; r}=h^{rs}(D(C)_mG_{ns}+D(C)_nG_{ms}-D(C)_sG_{mn})
\ee
We find
$$
L_{mn}^{\;\; r}=h^{rs}(\p_mG_{ns}+\p_nG_{ms}-\p_sG_{mn}-
2C_{mn}^{\;\; t}g_{ts}) -2SC_{mn}^{\;\; t} u_t h^{rs}A_s
$$
and
$$
L_{mn}^{\;\; r}=h^{rs}(\p_mG_{ns}+\p_nG_{ms}-\p_sG_{mn})-2C_{mn}^{\;\; r}
+2S C_{mn}^{\;\; t} u_t N^r.
$$
We may now ask for its $\veps$ gauge variation \Ref{SEC}:
$$
\delta L_{mn}^{\;\; r}=h^{rs}(\p_s\veps(u_m\p_nS+u_n\p_mS)-
\p_sS(u_m\p_n\veps+u_n\p_m\veps)).
$$
Again let us first consider the case of a covariantly constant Killing
vector ($S=1$), then
$L$ is at the same time a tensor and an ``$\veps$-tensor" and we find:
\be
C_{mn}^{\;\; r}-P_{mn}^{\;\; r}=-\frc12  L_{mn}^{\;\; r}
\ee
where we used once more the conservation of $u_m$.
Now it follows that the simplest choice for the connection $C$ would be the
$d$-dimensional part of the Levi-Civita connection $P$, i.e. $L=0$.
Conversely we have to prove that the latter preserves both $h$ and $u$. This
is the case when $S=1$. Clearly we cannot hope to fix the $K$ ambiguity
discussed in section 4.2, it is a dynamical question to optimise this
choice so as to simplify the $d$-dimensional equations of motion.
We recover the result of the Remark at the end of section 5.1.
We shall restore the arbitrariness of the scalar function S=W shortly by a
more geometrical argument, so let us proceed to study the quasi-Maxwell
degrees of freedom.

Taking into account the Weyl rescaling, we define
\be
\cA := n^m A_m - \frc12 h'^{m n} A_m A_n = -\frc12 SN^v  \;\;\; , \;\;\,
\cA^m := h'^{m n} A_n - n^m =-SN^m     \la{covA}
\ee
These fields are indeed invariant under \Ref{newsym1}, and furthermore
coincide with the components of the Maxwell field $A_m$ to
lowest order. Under $\veps$-gauge transformations, we have
\be
\d \cA^m = h'^{m n} \p_n \veps    \;\;\; , \;\;\;
\d \cA = - \cA^m \p_m \veps      \la{Max1}
\ee
The only $\veps$ invariant quantities (field strengths) that can be
constructed from $\cA$ and $\cA^m$ are found to be
\be
\cF^{m n} := \Dti^m \cA^n - \Dti^n \cA^m \;\;\;  , \;\;\;
\cF^m := \hti^{m n} \p_n \cA + \cA^n \Dti_n \cA^m  \la{Max2}
\ee
where $\Dti \equiv D (\Pti )$. Alas, a little algebra reveals
that both $\cF^{m n}$ and $\cF^m$ vanish identically! In fact,
after some thought we should not be too surprised at this result:
the vanishing of \Ref{Max2} is nothing but a fully (i.e. $\ISO (d-1)$ and
$\veps$) covariant version of the conditions \Ref{nDn} and
\Ref{Dn}.

In summary it seems impossible to extract some remnant of the
Maxwell degrees of freedom (that one would have expected to exist
on the basis of ordinary Kaluza Klein theory) in a completely
covariant fashion. This is in accordance
 with previous results on
Galilean covariant theories which we reviewed in section 4.2,
and lends credibility to the claim that the so-called Newton
(our Newton-Coriolis) condition is not sufficient to single out
purely gravitational effects coming from Einstein theory. What
is new in our treatment is that we have traced the ``disappearance"
of the Maxwell degrees of freedom to their apparent incompatibility
with the symmetries of the theory\footnote{To be sure, these degrees
of freedom have not really disappeared, as we have repeatedly
emphasized, but rather become part of gravity. This is also
suggested by the fact that for $d=3$ the gravitational sector
is apparently not a topological theory, unlike in ordinary
Kaluza Klein theory.}, and that with eq. \Ref{Max2} we have found a
completely covariant expression of this fact. Also, the inclusion of
the Kaluza Klein scalar (dilaton) $S$ is entirely new. The subtle
interplay between the equations of motion and the kinematic restrictions
that must be imposed on the gravitational
connection to recover a true Galilean situation was discussed in section
4.2. Let us stress that despite our title we have actually found a
generalized Galilean geometrodynamics with its unescapable Coriolis
or Maxwell effects.

In \ci{Du} a mysterious but simple formula was exhibited for the affine
connection for the case $W=S=1$. Let us show now that one can with hindsight
generalize it to our
situation. In the (local) fibration by the Killing orbits any tangent vector
fields $X' , Y'$
to the orbit space can be lifted, up to some ambiguities, to
vector fields $X , Y$ that commute with the Killing vector field $\xi$.
The covariant derivative upstairs $(X \cdot D)Y$
projects uniquely downstairs when $W$ is constant to
$(X' \cdot D')Y'$. This is the key remark that becomes applicable
in our more
general situation once we have noticed that the Weyl rescaling described
above amounts to a redefinition
\be
G_{MN}=WG'_{MN} \;\;\; , \;\;\;  G^{MN}=W^{-1} G'^{MN}.
\ee
Then it is clear that $\p_v$ remains a Killing vector of the rescaled metric
and that now
\be
\xi'_M=\p_Mu
\ee
and the connection formula \Ref{FINALS} follows. Actually this Weyl
rescaling reduces to the previous one only after a change of ``boost
gauge" (the $\R$ subgroup of section 3.2).

\section{Equations of motion and hidden symmetries}

Having identified the proper covariant objects, we are
now ready at last to give the complete equations of motion
obtained after the dimensional reduction with a null Killing vector
and to address the question of whether they can be derived from
an action.

\subsection{Connection coefficients and equations of motion}
We will now rewrite the full Einstein equations of motion for the null
Killing reduction. With the technology developed in the previous
sections this is most conveniently done in a ``$(d+1)$-covariant''
form and by use of intermediate indices where the equations take their
simplest form. The Einstein equations in $d+1$ dimensions read
\be
R_{\a\b} \equiv \H{\a}{M} \H{\b}{N} R_{MN} = 0    \la{RMN}
\ee
and must be supplemented by the reduction condition
$\xi^M \p_M \equiv 0$. Our conventions regarding the
Riemann tensor have been given in section 3.3. The full connection
prior to the Weyl rescaling has been given in \Ref{Jul}, or
equivalently in \Ref{fullG}. We now write out the
connection coefficients, taking into account the Weyl rescaling
and making the decomposition into $d$-dimensional
indices completely explicit. In this way, we get
\be
\G_{\m \n}^{\;\; \r} = \Pti_{\m \n }^{\; \; \r}
     + \d^\r_{(\m} \p_{\n )} s -
    \frc12 g'_{\m \n} h'^{\r \s} \p_\s s     \la{G1}
\ee
which is just \Ref{FINALS}, and
\be
\G_{\m \n}^{\;\; \ve} = S \Big(  - \frc12 n^\r {\cL}_\r g'_{\m \n}
    - \frc12 g'_{\m \n} n^\r \p_\r s + u_{(\m} A_{\n ) \r} n^\r
    - \frc12 A_{\m \n} \Big)      \la{G2}
\ee
\be
\G_{\m \ve}^{\;\; \n} = \G_{\ve \m}^{\;\; \n} =
  -\frc12 S^{-1} u_\m h'^{\n \r} \p_\r s        \la{G3}
\ee
\be
\G_{\m \ve}^{\;\; \ve} = -\frc12 \big( \d_\m^\n + u_\m n^\n \big)
       \p_\n s   \;\;\;  , \;\;\;
\G_{\ve \m}^{\;\; \ve} = +\frc12 \big( \d_\m^\n - u_\m n^\n \big)
       \p_\n s             \la{G4}
\ee
\be
\G_{\ve \ve}^{\;\; \m} = \G_{\ve \ve}^{\;\; \ve} = 0  \la{G5}
\ee
which shows explicitly that only the components of
$\G_{\a \b}^{\;\; \g}$ with $\g = \ve$ have an antisymmetric part.
For the $(d+1)$-dimensional trace of the connection, we obtain
\be
\G_{\a \m}^{\;\; \a} \equiv \G_{\n \m}^{\;\; \n} +
   \G_{\ve \m}^{\;\; \ve}  = \Pti_{\n \m}^{\;\; \n} +
  \frc12 (d+1) \p_\m s   \la{G6}
\ee
{}From these expressions, we can now obtain the corresponding ones
for vanishing scalar field $s$ (which we have not given so far)
by specializing to $s=0$.

\Ref{G1}--\Ref{G5} can now be substituted into \Ref{Ricci} to obtain the
equations of motion after some calculation. The $(\ve \ve)$-component
of \Ref{RMN} turns out to be identically satisfied:
\be
R_{\ve \ve} \equiv 0.
\ee
This is, of course, expected as we already used this equation as
an input to rewrite the Killing one-form in
\Ref{normality}. The remaining components of $R_{\a \b}$, however,
give rise to non-trivial equations. Discarding an overall factor,
we see that
\be
  \tilde D_\r \p_\s s + \frac{d-1}{2} \p_\r s \p_\s s  =0
\la{Rmuphi}
\ee
is just the scalar field equation (with $\Dti \equiv D (\Pti )$
as we already explained). As anticipated, it depends
on $s$ only through its derivatives. Furthermore, this equation
of motion involves the covariant {\sl transverse} Laplacian, and thus
can be regarded as a generalization of the transverse
Laplace equation obeyed by gravitational plane waves \ci{Bri}. When
$d=1$ the second term of the middle expression vanishes but the
two dimensional action is topological hence
\be
R_{\m \ve} \equiv \frc12 u_\m R.
\ee
Finally
\be
R_{\m \n} = \Rti_{\m \n} + \frac{d-1}{2} \bigg( \Dti_\m \p_\n s
       - \frc12 \p_\m s \p_\n s \bigg) =0  \la{Rmunu}
\ee
is Einstein's equation in $d$ dimensions, where $\Rti_{\m \n}  \equiv
R_{\m \n} (\Pti )$ and where we used the previous equation of motion and
restricted ourselves to the case $d\neq1$.
The fact that the Ricci tensor comes out to be symmetric is a useful
check on our calculations, because all antisymmetric contributions
arising at intermediate stages of the calculation must cancel out.
As we have repeatedly pointed out, the Maxwell field must
be absorbed into the connection to maintain covariance; consequently,
only the scalar ``matter" field can act as a source term in
\Ref{Rmunu}.

Observe also that this equation is more general
than \Ref{GalEinst} but that the dilaton decouples if one starts
in two dimensions as one might have expected; in that case one obtains
\be
R_{\m \n} \equiv \frc12 Sg'_{\m\n} R
\ee
\be
\Rti_{\m \n} = 0
\ee
and
\be
SR=h'^{\r \s} \tilde D_\r \p_\s s
\ee
is unconstrained.

We also note the following difference with ordinary (non-null)
Kaluza Klein theories. There the components $\Rti_{\ve \ve}$ and
$\Rti_{\m \ve}$ would have yielded the equations of motion for the
scalar and the Maxwell fields, respectively. Here, the first
equation is empty, while the second gives the scalar field equation
rather than Maxwell's equation. This is possible only because
of the presence of the covariantly constant vector $u_\m$, which has
no analog in the non-null case.

\subsection{An action}

The situation we found ourselves in seems as we noticed ill-adapted to the
construction of an action for two reasons. The first difficulty
arises from the fact that we have already used one of the
equations of motion, namely $R_{vv} =0$, as an input; it can be
surmounted by simply eliminating the corresponding component
$G^{vv}$ of the inverse metric.
However, this is not a covariant procedure, and we would
not expect the resulting action to be fully covariant either.
A second source of difficulties is the missing component
$G_{vv}$, which has been ``frozen" to zero.

Previous attempts to construct an action within the
purely Galilean covariant framework (i.e. in $d$ dimensions)
have encountered related difficulties (see \ci{Goe} for a
recent discussion), and so far no satisfactory action seems to be
known. One particular problem which arises in the $d$-dimensional
context is that
the covariant metric $g'_{\m \n}$ is not unique, it is degenerate and
thus has vanishing determinant. However the moving frame being conserved
up to a unimodular transformation there is an invariant density factor and
corresponding invariant antisymmetric tensor densities.
We have for the $d$-dimensional Weyl rescaled frame density:
\be
\p_\m {\rm log} \, e' \equiv
     \p_\m {\rm log} \, \EN' = \Pti_{\m\n}^{\;\; \n},
\ee
see also for example \ci{Di}
for the definition of the density factor without moving frames. On the
other hand, it was proposed in \ci{Du} to construct an action in
$d+1$ dimensions by introducing a Lagrange multiplier to enforce
the condition $\xi^M \xi_M =0$. This seems unjustified in view of the
extra constraint of the covariant constancy of the null vector, furthermore
the Lagrange multiplier remains undefined and one equation is still missing
after this manipulation.

We will here follow a somewhat
different route, also invoking the $(d+1)$-dimensional
ancestor theory, but avoiding the use of Lagrange multipliers.
An obvious argument in favour of starting from $d+1$ dimensions is
the existence of the non-degenerate metrics there (our $\Gnull_{\a \b}$
and $G_{MN}$). Taking into account the Weyl rescaling and the
presence of the dilaton, the density factor is
\be
E = \sqrt{G} = {\mkreis{E}}'(h', n)
        \exp \Big( \frac{d+1}{2} s \Big)                   \la{density}
\ee
Observe that it is independent of $A_\m$ as required by gauge
invariance, as well as invariant under \Ref{newsym1}.
It is equivalent to choose $h'$ and $n$ or $g'$ and $u$ as independent
variables.

The action density we propose is then essentially Einstein's action in
$d+1$ dimensions, written out in terms of intermediate indices, viz.
\be
{\cal L} = E \Gnull^{\a \b} R_{\a \b} =
   E \Big( \Gnull^{\m \n} R_{\m \n}   + 2 \Gnull^{\m \ve} R_{\m \ve}
     + \Gnull^{\ve \ve} R_{\ve \ve}  \Big)  \la{action1}
\ee
where we have given the last term only for the sake of clarity:
it actually vanishes because $\Gnull^{\ve \ve} = 0$ (or because
$R_{\ve \ve}=0$, see the foregoing section). Substituting the
expressions for the $(d+1)$-dimensional Ricci tensor and using
\Ref{density}, we obtain
\be
{\cal L} =
{\mkreis{E}}' \exp \Big( \frac{d-1}{2} s \Big) h'^{\m \n}
\bigg( \Rti_{\m \n} - \frac{d(d-1)}{4} \p_\m s \p_\n s \bigg)
\la{action2}
\ee

To verify that this is indeed the correct action density, we must now show
that the equations of motion \Ref{Rmuphi} and \Ref{Rmunu} follow from
\Ref{action2} by variation of the basic fields. For this, two crucial
points must be kept in mind. First of all, here we shall be using second
order formalism, i.e. we regard the connection
$\Pti_{\m \n}^{\;\; \r}$ as a dependent
field as explicitly defined by \Ref{Ptilde}. We shall maintain
zero torsion and hence the condition $\d u_\m = \p_\m  \d u$.
Secondly, in the space
of contravariant metrics $h'^{\m \n}$, the variations must be
performed in such a way that $h'^{\m \n}$ remains degenerate with
precisely one zero eigenvector.
They are therefore subject to the constraint
$\d h'^{\m \n} u_\n + h'^{\m \n} \d u_\n =0$.
Contracting with $u_\m$, we obtain
\be
\d h'^{\m \n} \, u_\m u_\n = 0      \la{dhuu}
\ee
Consequently the coefficient of $\d h'^{\m \n}$ in the variation of the
action will only be determined
up to terms of the form $\r u_\m u_\n$.

Variation of the dilaton $s$ yields the following equation
\be
h'^{\m \n} \bigg( d \Dti_\m \p_\n s + \frac{d(d-1)}{4} \p_\m s \p_\n s
+  \Rti_{\m \n} \bigg) =0   \la{var0}
\ee
Strictly speaking \Ref{var0} holds only for $d\neq1$ as we have
dropped a factor
$(d-1)/2$. To eliminate the Ricci tensor from \Ref{var0} and to arrive at
an equation involving $s$ alone, we must first analyze the
remaining equations obtained by varying the other fields.
Varying all fields except $s$, we get
\beql
\d_{{\rm grav}} {\cal L}  &=&
\EN' \exp \Big( \frac{d-1}{2} s \Big)  \d h'^{\m \n}
\Big( \Rti_{\m \n} - \frac{d(d-1)}{4} \p_\m s \p_\n s \Big)  \zeile
&+& \EN' \exp \Big( \frac{d-1}{2} s \Big) \big( \EN^{-1} \d \EN \big)
\, h'^{\m \n} \Big( \Rti_{\m \n} -
\frac{d(d-1)}{4} \p_\m s \p_\n s \Big)         \zeile
&+& \EN' \exp \Big( \frac{d-1}{2} s \Big)  h'^{\m \n}
\Big( \Dti_\m  \d \Pti_{\r \n}^{\;\; \r}
    - \Dti_\r \d \Pti_{\m \n}^{\;\; \r}   \Big)     \la{var1}
\eeql
where
\be
\EN'^{-1} \d \EN' = - \frc12 g'_{\m \n} \d h'^{\m \n}
   + n^\m \d u_\m                 \la{var2}
\ee
Upon partial integration, the third
line in \Ref{var1} becomes
\be
\Big( \frac{d-1}{2} \Big)\EN' \exp \Big( \frac{d-1}{2} s \Big) \p_\m s
\bigg( 2 \d \Pti_{\r \n}^{\;\; (\m} h'^{\n ) \r} -
 2 h'^{\m \n} \p_\n \big( \EN'^{-1} \d \EN'
 \big) \bigg)  \la{var3}
\ee
where we made use of $\d \Pti_{\n \m}^{\;\; \n} = \d ( \EN'^{-1} \p_\m
\EN' ) = \p_\m ( \EN'^{-1} \d \EN' )$. To further
evaluate \Ref{var3}, we will now use
\be
2 \d \Pti_{\r \n}^{\;\; (\m} h'^{\n ) \r} =
  - \Dti_\n \d h'^{\m \n}     \la{var4}
\ee
which is a consequence of the requirement that the covariant constancy
of $h'^{\m \n}$ be preserved under the variation along the space
of degenerate contravariant metrics. Note that all variations can
be parametrized in terms of only $\d s$, $\d h'^{\m \n}$
and $\EN'^{-1} \d \EN'$. The latter comprises the effect of the
variation of all the gravitational fields other than
$h'^{\m \n}$, according to \Ref{var2} only $\d n^\m$ appears, in particular
the variation of $\cal L$ does not depend on $\d A_\m$.

We pause here to point out that shifting $A_\m$
by $\d A_\m = u_\m f$ (such variations of $A_\m$ are
physical, contrary to the gauge shifts \Ref{newsym1}) changes
the Ricci tensor according to
\be
\d \Rti_{\m \n} \propto u_\m u_\n \Dti_\r (\hti^{\r \s} \p_\s f)
\ee
and hence does not change the action. Physically this means that its
equation of motion is left arbitrary by our $d$-dimensional
variational principle. Conversely we could hide the
arbitrariness of the equation of motion for $ R^{\ve\ve}$
that is due to \Ref{dhuu}, as mentioned above, by the appropriate
redefinition of the field $A_\m$.

Integrating the last line
of \Ref{var1} by parts once more and collecting terms,
we see that the terms multiplying $\d h'^{\m \n}$
combine precisely into the left hand side of \Ref{Rmunu}.
The terms multiplying
$\EN'^{-1} \d \EN'$ must then be combined with \Ref{var0}. After a
little reshuffling, these two equations are just the
scalar field equation \Ref{Rmuphi} and the trace of \Ref{Rmunu}.
Once more the variational principle for the reduced action leads to all
equations of motion but one.
Finally, we see again that the dilaton field decouples for $d=1$; this
just reflects the appearance of conformal symmetries in two dimensions.

\subsection{Hidden symmetries}
The Ehlers $SL(2,\R)$ duality transformations act on germs of solutions
of Einstein's vacuum equations admitting one non null Killing vector.
A $d+1=4$ covariant presentation is given in \ci{Ger} for the action of
the subgroup $SO(2)$. In a footnote Geroch
remarks that some action remains after a careful limiting procedure is
taken where the norm of the Killing vector tends to zero. We shall
develop this idea carefully in another paper but we may mention here
the following identity:
\be
G_{MN}= g_{MN}+2\xi_{(M} A_{N)}
\ee
where $g_{Mv} := 0$ and $A_v := 1$. Up to rescalings the Geroch action
amounts in the null case to the addition to the one-form $A_M$ of the
potential one-form whose exterior derivative is dual to the two-form
$dW\wedge du$. In the case of $pp$-waves it is easily found to be an
$\veps$-gauge transformation, in the case of say van Stockum solutions
it adds a constant ``electric" field and seems non trivial, in particular
it changes the symmetry properties.

\section{Conclusions}
This work suggests to investigate the addition of matter in order to
hunt for extra hidden symmetries in the case $d=3$. The addition of a true
Maxwell field is easy, it leads to the so-called
magnetic limit of electromagnetism \ci{LLB}. The addition of one or two
gravitinos should follow using standard techniques. This paper furnishes
all the required tools to permit the addition of fermionic fields.
 The massless sector of
closed string theory seems particularly interesting as already mentioned.
The role of the antiaxial gauge might deserve some more investigation.

We have worked out the $SO(2)$ action mentioned in \ci{Ger} contrary to
what is stated in the rest of the literature it does act {\sl nontrivially}
on the space of solutions admitting a null Killing vector.

As far as physics is concerned, we have been discussing the transverse
gravitational field seen by particles moving say along geodesics in such
backgrounds. Transverse meaning here that we consider the motion of the
projection on the Killing orbit space. Let us note a nice general
result. The scalar
product of the Killing vector and the velocity of such a particle is a
constant of the motion. In the case of a non null Killing vector it can
be interpreted as the electric charge. Here we obtain:
\be
\xi_M \frac{dx^M}{d\tau} = W \frac{du}{d\tau}
\ee
so we find - up to dilatonic effects we shall not discuss here - that
(for $W=1$) the absolute time, $u$, is an affine parameter for the geodesic
motion. The improvement of
the action principle and the study of constraints are prerequisites for a
quantization of that sector.

Finally it is amusing to speculate that quantum corrections will spoil the
classical equations of motion hence introduce torsion, but torsion is
known to be coupled to spin. Phrased differently the twist of a null
geodesic would be a natural manifestation of spin. We hope to return to some
of these issues later.

\bigskip

{\bf Acknowledgments:} This work was supported in part by Orsay
University, Deutsche Forschungsgemeinschaft and the EU human
capital and mobility program contract ERBCHRXCT920069. We would like
to thank the II. Institute of Theoretical Physics, University of
Hamburg, and the ENS, Paris, respectively, for hospitality,
and T. Damour for some useful references.

\end{document}